\documentstyle[12pt,aaspp4,amstex]{article}

\def\ltsima{$\; \buildrel < \over \sim \;$}
\def\simlt{\lower.5ex\hbox{\ltsima}}
\def\gtsima{$\; \buildrel > \over \sim \;$}
\def\simgt{\lower.5ex\hbox{\gtsima}}

\begin{document}

\title{Cosmic Reionisation by Stellar Sources: Population III Stars}

\author{Aaron Sokasian\altaffilmark{1}, Naoki Yoshida\altaffilmark{1,2},
Tom Abel\altaffilmark{3}, Lars Hernquist\altaffilmark{1}, \\ and Volker
Springel\altaffilmark{4}}

\altaffiltext{1}{Harvard-Smithsonian Center for Astrophysics, 
60 Garden Street, Cambridge, MA 02138, USA}
\altaffiltext{2}{National Astronomical Observatory of Japan,
Mitaka, Tokyo 181-8588, Japan}  
\altaffiltext{3}{Department of Astronomy and Astrophysics, 
Penn State University, 525 Davey Lab,
University Park, PA 16802, USA} 
\altaffiltext{4}{Max-Plank-Institut f$\ddot{\text{u}}$r Astrophysik, 
Karl-Schwarzschild-Stra$\beta$e, 1,
85740 Garching bei M$\ddot{\text{u}}$nchen, Germany}

\authoremail{asokasia@cfa.harvard.edu}

\begin{abstract}

We combine fast radiative transfer calculations with high resolution
hydrodynamical simulations to study an epoch of early hydrogen
reionisation by primordial stellar sources at redshifts $15\lesssim z
\lesssim 30$.  We consider the implications of various
local and global feedback mechanisms using a set of models which
bracket the severity of these effects to determine, qualitatively, how
they may have influenced the global star formation rate and the details
of hydrogen reionisation.  With relatively conservative assumptions,
most of our models suggest that population III star formation proceeds
in a self-regulated manner both locally and globally and, for a
conventional $\Lambda$CDM cosmology, can significantly reionise the
intergalactic medium between $15\lesssim z \lesssim 20$ as long as a
large fraction of ionising photons can escape from these earliest
galaxies.

We then combine these results with our earlier work focusing on the
role of population II stars in galaxies with virial temperatures
$\simgt 10^4$\,K at redshifts $5\lesssim z\lesssim 20$. Hence, we
construct a complete reionisation history of the Universe which
matches the Thomson optical depths as measured by the WMAP satellite
as well as the evolution of the Gunn Peterson optical depth as seen in
the absorption spectra of the higest redshift quasars.  We find that
even with conservative estimates for the impact of negative feedback
mechanisms, primordial stellar sources contribute significantly to
early reionisation. Future observations of a Thomson optical depth of
$\tau_e\simgt0.13$ would bolster the claim for the existence of
population III stars similar to the ones studied here.


\end{abstract}


\section{INTRODUCTION}

The reionisation history of the intergalactic medium (IGM) holds many
important clues about the onset of structure formation and the nature
of the first luminous sources.  Advances made in the past few years
are making it possible to directly probe this history.  Spectroscopic
observations of distant quasars by Fan et al. (2000) and Djorgovski et
al. (2001) revealed that the IGM was highly ionised at $z\simeq6$.
Subsequent high resolution studies of three $z>5.8$ quasars by Becker
et al. (2001) showed a significant increase in Ly$\alpha$ absorption
from redshift 5.5 to 6.0 with the first possible detection of a
Gunn-Peterson trough in the spectrum of the $z=6.28$ quasar SDSS
1030.10+0524. More recently, White et al. (2003) presented an
additional high resolution spectrum of a quasar at $z=6.37$ (Fan et
al. 2003) which also appears to have an extended Gunn-Peterson trough.
Together, these results indicate that the Universe was reionised only
a short time earlier, implying a relatively late epoch for complete
reionisation.

Results from the Wilkinson Microwave Anisotropy Probe (WMAP) satellite
(Kogut et. al 2003; Spergel et al. 2003) have challenged this scenario
by suggesting that much of the IGM was ionised earlier than inferred
from the SDSS quasars.  In particular, the measured correlation
between polarisation and temperature yields an electron optical depth
to the CMB surface of last scattering of $\tau_e=0.17$, corresponding
to an instantaneous reionisation epoch $z_r=17$.  The uncertainty
associated with this measurement depends on fitting all parameters
concerned with the TT power spectrum and the TE cross power
spectrum. After a careful analysis, Kogut et al. (2003) have reported
a 68\% confidence range of $0.13 < \tau_e < 0.21$ corresponding to an
instantaneous reionisation epoch $14<z_r<20$.  In contrast, if
reionisation was completed around $z\sim 8-10$, as is suggested by the
SDSS quasars, the electron optical depth would be only $\tau_e\sim
0.05-0.06$.  Although the discrepancy is only at the $\sim 3 \sigma$
level, it is timely to explore the connection between the two sets of
observations.

In a previous paper (Sokasian et al. 2003; hereafter Paper I), we used
a high resolution cosmological simulation in conjunction with a fast
3D radiative transfer code to study how stellar sources similar to
those seen in local galaxies, i.e. population II type, contributed to
the reionisation history of the IGM in the redshift interval
$6\lesssim z \lesssim 20$.  This work was similar to other related
numerical studies (e.g., Gnedin 2000; Ciardi et al. 2000; Razoumov et
al. 2002; Ciardi et al. 2003) except that source intensities were
derived directly from intrinsic star formation rates computed in the
underlying hydrodynamical simulations. In addition, the high mass
resolution of the simulation allowed us to include sources down to
$M\sim4.0 \times 10^{7}\ h^{-1}$ M$_{\odot}$ in a $10.0 \ h^{-1}$ Mpc
comoving box, which proved to have a significant impact on the
reionisation process.  Our results showed that the star-formation
history inferred by Springel \& Hernquist (2003) based on detailed
hydrodynamic simulations and estimated analytically by Hernquist \&
Springel (2003) led to a relatively late reionisation epoch near
$z\simeq7-8$ for a plausible range of escape fractions. Additionally,
we generated synthetic spectra from a range of models with different
escape fractions and showed that values in the range 10-20\% led to
good statistical agreement with observational constraints on the
neutral fraction of hydrogen at $z\sim6$ derived from the $z=6.28$
SDSS quasar of Becker et al. (2001). However, the relatively late
reionisation epoch predicted by these models cannot account for the
high electron scattering optical depth inferred from the WMAP
measurements.

In an effort to study how this difference could be reconciled, we
employed heuristic models with an evolving ionising boosting factor and found
that in order to simultaneously match the WMAP and SDSS constraints,
one requires an $f_{\rm esc}=0.20$ model with a boosting factor that rises
from unity at $z\simeq6$ to $\gtrsim50$ near $z\sim18$.  In this
picture, large boosting factors mean that the stellar production rate
of ionising photons is much higher than that which would be produced
according to our model assumptions.  One mechanism that would boost
the production rate of ionising photons would be if the stellar IMF
evolved so that it became increasingly top-heavy at high redshift.
While possible, this scenario would still result in an IGM that is
highly ionised at $z<13$, a condition which may be inconsistent with
the thermal history of the IGM. More specifically, Hui \& Haiman
(2003; see also Theuns et al. 2002) have shown that as long as the
Universe is reionised before $z=10$ and remains ionised thereafter,
the IGM would reach an asymptotic thermal state which is too cold
compared to the temperature inferred from Ly$\alpha$ forest
observations at $z\sim 2-4$. This conclusion, however, relies on
assumptions regarding the ionising spectrum. For example, if the
ionising spectrum is harder than the typical quasar spectrum above the
H {\small I} and He {\small I} ionisation thresholds, then order unity
changes in the ionisation fraction of H {\small I} and He {\small I}
are necessary in the range $6<z<10$ in order to sufficiently heat up
the IGM within the observation constraint at $z\approx 4$. However, if
the spectrum is relatively soft (more typical of population II stars)
then this condition is avoided as long as He {\small II} reionisation
occurs relatively late near $z\sim4$ (see, for example, Sokasian
et al. 2002).

Another difficulty with boosting the production rate of ionising
photons, as emphasised by e.g. Yoshida et al. (2003a,e) is that the IMF
would need to remain top heavy for many generations of high mass
stars.  These stars have typical lifetimes $\sim$ a few million years.
However, the total elapsed time between $z=18$ and $z=10$ is more than
a hundred times longer; therefore, it would be necessary for massive
star formation to proceed efficiently for more than 100 generations.
Studies by e.g. Omukai (2000), Bromm et al. (2001), and Schneider
et al. (2002) indicate that very massive stars can form
only if the metallicity of the gas lies below a critical threshold,
$Z \simlt 10^{-3.5} Z_\odot$.  It appears problematic for the gas in
star-forming halos to remain sufficiently chemically pristine for
massive star formation to continue for over 100 stellar generations.

The combined observational constraints summarised above suggest that
the reionisation history of the Universe was complex, perhaps having
had multiple reionisation epochs. Such a scenario has been considered
by Cen (2003b) and Wyithe \& Loeb (2003b) who pointed out the
possibility that the Universe was first reionised by an early
generation of massive, metal-free (population III) stars, but that
then much of the IGM recombined once these stars were no longer able
to form, and the Universe was subsequently reionised a second time by
the next generation of population II stars. In hierarchical models of
structure formation, these ``first (metal-free) stars'' are thought to
form through molecular hydrogen (H$_2$) cooling in halos with masses
larger than $\sim7 \times 10^5$ M$_{\odot}$ at redshifts as high as
$z\sim 20 - 30$, provided that the halos are dynamically cold and
contain sufficiently larger numbers of hydrogen molecules (Yoshida
et al. 2003c) .  

Recent progress in simulating the formation of the very first objects
in the universe (Abel et al. 1998; Bromm, Coppi \& Larson 1999) has
shown that the first luminous objects to form in CDM models are stars.
In addition, because the first H {\small II} regions are capable of
driving out most of the gas from the halo, it is unlikely that massive
seed black holes are able to accrete within a Hubble time after they
formed (Whalen, Abel, \& Norman 2003). The notion that these
population III stars played a role in cosmological reionisation in
Cold Dark Matter (CDM) models has been discussed widely (e.g.,
Couchman \& Rees 1986; Barkana \& Loeb 2001; Mackey, Bromm, \&
Hernquist 2003; Venkatesan, Tumlinson \& Shull 2003; Cen 2003a; Wyithe
\& Loeb 2003a; Yoshida et al. 2003a,b).  Simulations indicate that
these stars which form in ``mini-halos'' should be massive $\gtrsim
100$ M$_{\odot}$ (Abel, Bryan, \& Norman 2000, 2002; Bromm, Coppi \&
Larson 2002). Whereas Abel et al. (2002) predict one massive star per
halo Bromm et al. (2002) suggest that multiple
stars can be produced in each dark matter
halo. We discuss both scenarios in this paper. Such population III
stars are efficient UV emitters because of their approximately
constant effective surface temperatures near $10^{5}$ K with lifetimes
of $\sim3\times10^6$ yr, independent of their mass (e.g. Bromm,
Kudritzki \& Loeb 2001; Schaerer 2002 for recent discussions).  In
fact, these stars are capable of producing up to $\sim30$ times more
ionising photons per baryon compared to population II stars with a
Salpeter IMF.

Although the impressive yield of ionising photons from population III
stars appears promising in terms of a possible solution to the early
reionisation epoch suggested by WMAP, there are still many questions
remaining as to whether enough of these first generation stars
could have formed to cause cosmic reionisation. In general,
semi-analytic modeling has been used to address these questions (see,
e.g., Wyithe \& Loeb 2003a,b; Cen 2003a,b); however, the predictions
of these models are often uncertain because of the relatively crude
assumptions used to relate luminous objects to the dark matter halos
which host them.  For example, Yoshida et al. (2003c) have shown
that simple criteria based on only the mass or virial temperatures of
halos can fail by an order of magnitude or more because of dynamical heating
by mergers.  Additionally, the effects of feedback processes are very
difficult to assess in semi-analytic models. The UV radiation in the
Lyman-Werner (LW) band (11.26-13.6 eV) from these massive stars can
photo-dissociate the fragile H$_2$ molecules which are responsible for
their formation. This negative feedback can significantly limit
population III star formation and thereby substantially reduce the
effective ionising flux in the early Universe. Conversely, it has been
argued that feedback from X-rays can promote H$_2$ production by
boosting the free electron fraction in distant regions (Haiman, Rees
\& Loeb 1996; Oh 2001). Efforts to explore feedback effects on cooling
and collapsing gas have been pursued by many groups (e.g., Dekel \&
Rees 1987; Haiman, Rees, \& Loeb 1997; Haiman, Abel, \& Rees
2000). However, these studies have primarily relied on semi-analytic
methods which assume spherically symmetry and ignore
details regarding formation dynamics and locations of halos in complex
density fields.

A more realistic analysis of the problem requires detailed numerical
simulations capable of capturing the physics of early structure
formation and related feedback effects.  This was first attempted by
Machacek, Bryan, \& Abel (2001; see also Machacek, Bryan, \& Abel
2003) who employed an Eulerian adaptive mesh refinement simulation to
investigate the quantitative effects of a background radiation field
on the subsequent cooling and collapse of high-redshift pre-galactic
clouds. Ricotti et al. (2002a,b) followed up on the numerical approach
by including radiative transfer effects and found that the formation
of small halo objects is not significantly suppressed by the
dissociating background. More recently, Yoshida et al. (2003c) used
high resolution cosmological simulations which included a careful
treatment of the physics of chemically pristine gas and the impact of
a dissociating UV background to study the statistical properties of
primordial halos in cosmological volumes. In a subsequent study,
Yoshida et al. (2003b) coupled radiative transfer calculations to
their analysis to investigate reionisation by an
early generation of stars. More specifically, they identified
plausible sites of star formation in their cosmological volume where
they then placed massive stars with $M = 300$ M$_{\odot}$ under the
usual ``one star per halo'' assumption. They ran the multi-source
radiative transfer code described in Paper I on a $200^3$ grid with
the assumption that source lifetimes are 3 million years and that the
escape fraction for ionising radiation from the halos is unity. To
mimic the strong radiative feedback in H {\small II} regions, they
implemented a ``volume exclusion effect'' which disabled sources
within already ionised regions. The results from their analysis
indicated that in the standard $\Lambda$CDM cosmology, a sufficient
number of population III sources turn on to fully reionise the volume
by $z\simeq18$. By merging the subsequent ionisation history from
population II sources (adopted from Paper I), they were able to
compute a total optical depth to Thomson scattering of $\tau_e\sim
0.14$, in reasonable agreement with the WMAP result.

Encouraged by the results from Yoshida et al. (2003b), the aim of this
paper is to use the numerical approach outlined in Paper I to perform
a more rigorous study of the parameter space associated with the
ability of population III stars to reionise the Universe. In
particular, by tracking individual clouds of dense molecular gas
directly, we are able to implement simulated feedback effects
explicitly. We explore a number of plausible models in which we
incorporate feedback effects in an {\it ad hoc} fashion and examine
the corresponding ionisation history of the Universe. By coupling this
history with the results from the population II analysis conducted in
Paper I, we are able to assess the overall success of each model in
terms of the WMAP and SDSS observations.

This paper is organised in the following manner. In \S 2 we describe
the cosmological simulation used in our analysis. Then in \S 3 we
describe our methodology for source selection and present a detailed
discussion of how we approximate related feedback effects.  The
results of the simulation are presented in \S 4. Included in this
section is a discussion related to the reionisation history of the
Universe from the combined effect of population II and population III
stars. Finally in \S 5, we provide a summary and state our
conclusions.

\section{UNDERLYING COSMOLOGICAL SIMULATION}

The underlying simulation used in this paper was performed using the
parallel Tree-PM/SPH solver GADGET2, in its fully conservative entropy
form (see Springel \& Hernquist 2002). The simulation follows the
non-equilibrium reactions of nine chemical species (e$^-$, H, H$^+$,
He, He$^+$, He$^{++}$, H$_2$, H$^+_2$, H$^-$) using the reaction
coefficients compiled by Abel et al. (1997). The cooling rate of Galli
\& Palla (1998) is employed for molecular hydrogen. In this paper, we
will focus on a $\Lambda$CDM model with matter density $\Omega_m=0.3$,
baryon density $\Omega_b=0.04$, cosmological constant
$\Omega_{\Lambda}=0.7$ and expansion rate at the present time $H_o=70$
km s$^{-1}$ Mpc$^{-1}$. The spectral index of the primordial power
spectrum has been set to $n_s=1$ and we normalise the fluctuation
amplitude by setting $\sigma_8=0.9$. Further details of the simulation
and methods for generating initial conditions are given in Yoshida et
al. (2003c) and Yoshida et al. (2003d), respectively.

The particular simulation we use employs $2\times324^3$ particles in
a cosmological volume of 1 Mpc on a side, corresponding to a mass
resolution of 160 M$_{\odot}$ and 1040 M$_{\odot}$ for gas and dark
matter particles, respectively. Convergence tests using higher
resolution simulations indicate that the mass resolution adopted here
is sufficient to follow the cooling and collapse of primordial gas
within low mass ($\sim 10^6$M$_{\odot}$) halos.

Owing to the relatively small size of our simulation box, the
fundamental fluctuation mode starts to become non-linear at
$z\approx16$, forcing us to stop our analysis at this redshift. We
discuss the limitations of such a small volume in the context of
predicting cosmic reionisation in \S 4.

\section{SOURCE DEFINITION AND FEEDBACK EFFECTS}

\subsection{Source Definition}

Our approach for defining sources relies on identifying virialised
halos which contain a reservoir of cold and dense gas that undergoes
star formation. This requires first locating dark matter halos
via a friends-of-friends (FOF) algorithm with a linking parameter
$b=0.164$ in units of mean particle separation, and discarding groups
with fewer than 100 dark matter particles. For each FOF group, we
compute a virial radius $R_{\rm vir}$ defined as the radius of the sphere
centered on the most bound particle of the group having an overdensity
of 180 relative to the critical density. Within this radius we then
measure the mass of gas which is cold ($T<500$ K) and dense
($n_{\text{H}}>500$ cm$^{-3}$). The clouds of molecular gas in these
reservoirs is expected to rapidly cool and exceed the characteristic
Jeans mass $M_J\sim 3000$ M$_{\odot}$ for the typical temperature
$T\sim200$ K and density $n_{\text{H}}\sim 10^3$ cm$^{-3}$ of the
condensed gas. The halos which contain these reservoirs of cold gas
are therefore considered to be sites of active star formation and
serve as potential locations for population III sources. In the top
panel of Figure 1 we show how cold gas accumulates in star-forming
halos as a function of redshift.  Here, star-forming halos ({\it solid
circles}) are plotted at each simulation snapshot (separated in time
by $3\times 10^6$ years) and serve to illustrate how cold gas
accumulates within individual halos over time.  In the bottom panel of
the same figure we show the total number of star forming halos as
function of redshift in our simulation volume. It is important to
point out that in carrying out our analysis, we construct a halo
merger history and use it to carefully track the star-forming gas that
is exchanged between the progenitors and descendants of each halo.

For each potential source site, we compute the amount of gas that
actually forms stars by multiplying the total amount of cold gas in
the halo by the free parameter $E_{\rm SF}$. Here $E_{\rm SF}$ can be
thought of as a universal star formation efficiency of {\it
dense-cold} gas as defined above to form sources and should not be
confused with similar definitions of star formation efficiencies based
on the {\it total} amount of gas in the halo (see, e.g. Cen
2003a,b). In this study, we consider two different values for this
parameter: $E_{\rm SF}=0.02$, and $E_{\rm SF}=0.005$. For a given
value of $E_{\rm SF}$, we apply a further restriction on the allowable
mass of stars $M_{\star}$ which can actually form.  More specifically,
we assume that the primordial IMF is moderately top-heavy and yields
stars with masses in the range 100-300 M$_{\odot}$.  We therefore
first discard all halos with star masses below $100$
M$_{\odot}$. Above this limit, we allow a range of star masses to be
incorporated into the halo up to a maximum of 900 M$_{\odot}$ under
the assumption that only a few stars can form at coincidentally close
times and thus survive the intense dissociating radiation from the
first star in the halo (see Omukai \& Nishi 1999; Abel, Bryan \&
Norman 2002) and its subsequent explosion (e.g. Bromm et al. 2003).  
Our assumption may therefore be considered as a
slightly more relaxed version of the ``one-star-per-halo''
prescription employed in related studies of this topic (see, e.g., Oh,
Nollett, Madau, Wasserburg 2001). Nevertheless, it should be noted
that the associated parameter space which will be explored in this
study leads to relatively few multi-source halos ($M_{\star}>300$
M$_{\odot}$) under the assumption of a moderately top-heavy IMF and
therefore leads to similar results.

Because of their high mass, population III stars are dominated by
radiation pressure and have luminosities close to the Eddington limit
$L_{\rm Edd}$.  This leads to a constant source lifetime estimated to be
$t\simeq 0.007Mc^2/L_{\rm Edd} \sim 3 \times 10^6$ yr. We use this value
as the universal lifetime for all our stars and also adopt it as the
time step over which we evolve our simulation. To compute ionising
intensities we first note the results presented in Bromm, Kudritzki,
\& Loeb (2001) who demonstrated that population III stars with masses
$\geq 300$ M$_{\odot}$ have a generic mass-scaled spectral form that
is almost independent of mass resulting in H {\small I} and He {\small
II} ionising photon production rates $1.6\times10^{48}$ s$^{-1}$
M$_{\odot}^{-1}$ and $3.8\times10^{47}$ s$^{-1}$ M$_{\odot}^{-1}$,
respectively. For 100 M$_{\odot}$ stars, the number of ionising
photons per solar mass is reduced by a factor of $\sim 2$ for H
{\small I} and $\sim 4$ for He {\small II} (Bromm et al. 2001;
Tumlinson \& Shull 2000). Given our assumption of the 100 - 300
M$_{\odot}$ mass range for individual stars, we compute ionisation
rates for halos by interpolating between the values quoted above for
the 100 M$_{\odot}$ and $\geq 300$ M$_{\odot}$ cases\footnote{Note
that in our simulations we carry out radiative transfer
calculations to track H {\small II} and He {\small III} regions only,
with the implicit assumption that for massive stars, the boundaries of
the ionisation zones for He {\small II} and H {\small II} coincide.
This serves to significantly speed up our calculations by allowing the
tracking of both zones via a single ray tracing calculation in H
{\small II}; see \S 2.1 of Paper I for further details.}. In the
relatively few halos where the total source mass exceeds 300
M$_{\odot}$, we adopt the median values of $1.2\times10^{48}$ s$^{-1}$
M$_{\odot}^{-1}$ for H {\small I} and $2.4\times10^{47}$ s$^{-1}$
M$_{\odot}^{-1}$ for He {\small II} under the assumption that these
halos contain multiple stars with masses in the range 100 - 300
M$_{\odot}$.

The quantity of ionising flux that can actually escape a halo and
enter into the IGM is parameterised by the escape fraction $f_{\rm
esc}$.  In this context, $f_{\rm esc}$ is defined to be the fraction
of ionising photons intrinsic to each source that escape the virial
radius of the halo and participate in the radiative transfer
calculations. It must be noted here that this parameter inevitably
also carries with it any uncertainties associated with the shape of
the galaxy, the gas and stellar density profiles, and the IMF and star
formation efficiency. At low redshifts, observations deduced from
$z<3$ starburst galaxies (see Heckman et al 2001; Hurwitz et al. 1997;
Leitherer 1995) suggest small values for $f_{\rm esc}$ around
$f_{\rm esc}\lesssim0.10$. However, it is important to note that these
observational results may be underestimating the true values owing to
undetected absorption from interstellar components. In fact,
observations of 29 Lyman break galaxies at $z\sim3.4$ by Steidel,
Pettini, Adelberger (2001) appear to suggest much larger escape
fractions. Furthermore, it is unclear how the relevant factors
associated with the escape fraction vary with redshift. Theoretical
work on $f_{\rm esc}$ at high redshifts (Wood \& Loeb 2000, Ricotti
\& Shull 2000) finds small values, although these estimates are highly
uncertain owing to the lack of information regarding the IMF and star
formation efficiencies at early times. In addition, the effect of
dust, complex gas inhomogeneity and gas dynamics, all of which may
strongly influence escape fractions, are not included in these
theoretical studies. More recently, Whalen et al. (2003) have studied
the ionisation environment of the first luminous objects using
detailed one--dimensional radiation hydrodynamical simulations and
find that the transition from D to R--type fronts occurs within only
$\sim 100,000$\,yrs after the star reaches the zero age main
sequence. Because these I--fronts exit the halo on timescales much
shorter that the stars' main sequence lifetime, their host halos can
have UV escape fractions of $\gtrsim 0.95$.

Given these recent results we employ relatively large escape
fractions, $f_{\rm esc}=1$ and $f_{\rm esc}=0.3$, in all the models we
consider.  The escape fraction of unity is motivated by the extreme
cases from Whalen et al. (2003) while the escape fraction of 0.3 is
meant to represent an alternate case where the surrounding environment
is much more complex and restrictive to ionising break-throughs.

\subsection{Feedback effects}

Feedback effects from the first stars inevitably exert prompt and
significant impact on subsequent star formation. Locally, population
III stars are expected to significantly affect their environments when
they explode as supernovae (e.g. Bromm et al. 2003).
This ``internal'' feedback will perturb
the gas and effectively regulate subsequent star formation in the
pregalactic cloud. Feedback from the first stars can also affect
structure formation on a global scale.  This is because the UV photons
below 13.6 eV produced by these stars can easily propagate through the
Universe and build up a soft UV background that can alter the
chemistry of distant regions. More specifically, studies have
suggested that even a feeble UV background can photo-dissociate the
fragile H$_2$ molecules necessary for cooling of the primordial gas
(see, e.g., Haiman, Rees, \& Loeb 1997; Haiman, Abel, Rees 2000;
Ciardi, Ferrara, \& Abel 2000; Machacek, Bryan, \& Abel
2001). Conversely, it has been argued that feedback may actually
enhance H$_2$ formation ahead of of the ionisation front (Ricotti,
Gnedin, \& Shull 2001) or through the boosting of the electron
fraction in dense regions by X-rays emitted from, for example, distant
supernovae remnants (Haiman, Rees \& Loeb 1996; Cen 2003a; but see
also Machacek, Bryan, \& Abel 2003 who find that the net effect of
X-rays is rather mild).

Presently, the technical difficulties involved with the incorporation
of radiative transfer calculations directly into cosmological
simulations prevent us from studying these effects numerically in a
fully consistent manner. We therefore adopt a more {\it ad hoc}
approach which attempts to include feedback effects through the
incorporation of simple, physically motivated criteria which
dynamically constrain the amount of stellar mass that forms. In the
following two sections we will describe our prescription for
parameterising internal and external feedback effects within the
context of our simulation.

\subsubsection{Simulating internal feedback effects}

Stars with masses in the range 140 M$_{\odot} \lesssim M \lesssim 260$
M$_{\odot}$ are expected to die as supernovae which release up to
$10^{53}$ ergs of kinetic energy into their surroundings (e.g. Heger
\& Woosley 2002). Since the gravitational binding energy for gas in
sub-galactic halos is $E_b\sim 10^{49} [M/(5\times 10^5)]^{5/3}
[(1+z)/20]$, these explosions can severely disrupt their immediate
surroundings, possibly {\it driving away} a large portion of the gas
from the halo (see, e.g., Bromm, Yoshida, \& Hernquist 2003). To mimic
this effect, we introduce an ``internal disruption factor'' $D_f$
which will parameterise the amount of gas that gets blown out. More
specifically, for every halo which turns on with star mass
$M_{\star}$, we flag a corresponding amount of star forming gas mass
$M_{\rm SF}$ (cold gas) equivalent to $D_f \times M_{\star}$ within that
halo as unusable for future star formation. The prescription for
calculating a new star mass for a given halo is then quantified at
each time step $i$ in terms of $E_{\rm SF}$ by the expression,
\begin{equation}
M_{\star}^i = E_{\rm SF}[M_{\rm SF}^{i-1} (1- E_{\rm SF}D_f) +\Delta M_{\rm SF}^i],
\end{equation}
where $M_{\rm SF}^{i-1}$ represents the amount of {\it usable}
star-forming gas mass from the previous step and $\Delta M_{\rm SF}^i$
is the amount of new star-forming gas accumulating in the halo at the
current step.  This prescription requires us to explicitly track the
history of gas in each halo directly, accounting for the accumulation
of new star-forming gas by continuing accretion from surrounding gas,
and also from mergers with other star-forming halos. In this analysis,
we explore a case where the disruption factor is small, $D_f=40$, and
also a case where the disruption factor is large, $D_f=120$. We admit
that these choices for the disruption factor are somewhat arbitrary.
However, our aim here is to use $D_f$ in conjunction with $E_{\rm SF}$
to create a range of different models which span the plausible
parameter space associated with internal feedback effects. In
particular, our choice of $D_f=40$ will describe a scenario where
clumps of condensed gas in the halo only inefficiently absorb the
energy released from a local supernova (possibly owing to a spatially
clustered distribution with a small cross section and large column
density to the explosion site) and are therefore marginally disrupted
($E_{\rm SF}\times D_f = 0.2 - 0.8$).  Such a low disruption factor
may also be motivated by the potential positive feedback which can
arise from propagating compression shocks induced by the exploding
star which may accelerate star formation within the halo (Mackey,
Bromm, \& Hernquist 2003; Salvaterra, Ferrara \& Schneider 2003). On
the other hand, our choice of $D_f=120$ will simulate the case where
energy absorption is efficient and effectively disperses a large
amount ($E_{\rm SF}\times D_f = 0.6 - 2.4$) of the gas. Here, values
for $E_{\rm SF}\times D_f$ close to unity correspond to the total
disruption of star-forming gas within the halo, while values larger
than unity are meant to represent an extreme disruption of the gas
extending beyond the halo. Formally, the latter scenario results in
negative star masses for some of the halos immediately after a
star-forming episode. In these cases we simply set the star mass to
zero, but continue to track the amount of {\it negative} star forming
gas mass in the halo, which decreases in time as new gas is
accumulated.  Negative star forming gas, in this context, represents a
ramification of strong internal feedback effects which not only
disrupt the available star forming gas in the halo but also inhibit
the future accretion of neighbouring gas surrounding the halo.  In all
cases, it is important to keep in mind that our parameterisation
of the disruption factor $D_f$ also carries with it the uncertainty
associated with whether or not the stars formed actually end their
lives as supernovae.

In Figure 2, we plot all the possible star-forming halos in our
simulation volume for the four different combinations of $D_f$ and
$E_{\rm SF}$ we tried.  Here, the two dotted horizontal lines
delineate the range of allowable star masses in halos ({\it solid
circles}) as described earlier.  In particular, halos with source
masses falling short of $100$ M$_{\odot}$ ({\it crosses}) are excluded
while those above the $900$ M$_{\odot}$ limit ({\it open-circles}) are
included, but with star masses artificially set to the upper limit of
$900$ M$_{\odot}$.  A comparison between this figure and the top panel
in Figure 1 clearly demonstrates the effect of including a disruption
mechanism in the halos. Namely, once halos acquire enough star mass to
turn on, the surrounding star forming mass in that halo is disrupted
to varying degrees and does not monotonically increase as a function of
time.  Consequently, a halo which has turned on at a given time may not
have enough undisrupted star-forming gas at subsequent time steps to
turn on again. This effect is especially noticeable in the right-hand
column where the disruption factor is high with the most dramatic case
indicated in the top-right panel. Here, the combination with a high
efficiency factor renders most halos incapable of turning on more than
once.  In the left-hand column, the disruption factor is 3 times
smaller and most halos are capable of turning on repeatedly. This is
especially true in the bottom-left panel where the efficiency is also
small and individual halos can be seen switching on continuously once
they surpass the minimum star mass threshold. Interestingly, in all
cases the most massive halo in our simulation volume appears to
contain enough star-forming mass at redshifts below 20 that the
disruption caused by a star mass of 900 M$_{\odot}$ switching on at
each step does not inhibit subsequent star formation. This seems like
a plausible scenario for halos with very deep potential wells that are
especially conducive to star formation. It must be noted that at some
level there exists a degeneracy between the effects produced by the
parameters $D_f$ and $E_{\rm SF}$. However, as Figure 2 demonstrates,
there are distinct differences between the two parameters in terms of
simulating the physical consequences of star formation, especially in
extreme cases. We therefore continue our analysis with this
distinction.

\subsubsection{Simulating external feedback effects}

We now turn to the issue of simulating external feedback effects which
have a more global impact. The first type of external feedback we
consider is related to the ``sphere of influence'' associated with the
ionisation zone surrounding a source. The gas within this region is
expected to be substantially photo-heated to temperatures in excess of
$2\times 10^4$ K when ionised by radiation with a spectrum typical of
metal-free stars (see Hui \& Haiman 2003). For example, a
$7\times10^{5}$ M$_{\odot}$ halo at $z\simeq20$ has a binding energy
of roughly $\sim 10^{49}$ ergs, which is considerably less that the
total ionising energy produced by a single massive star of order
$\sim1 \times 10^{54} (M_{star} / 200$~M$_{\odot})$ erg.  As a result,
star-forming gas in shallow potential wells located within these
ionised regions photo-evaporates and fails to produce stars. The
magnitude of the effect is difficult to gauge without a self
consistent treatment of radiative photo-heating effects linked
directly to the hydrodynamics of the gas. Since the radiative transfer
and gas dynamics are not coupled in our calculations,
we can only simulate the effect in an {\it ad hoc} manner. In
particular, we would like to qualitatively mimic the inhibiting
effects that photo-heating may have on star formation in halos as
function of the depth of their potential wells, or masses,
as seen in the simulations of e.g. Bromm et al. (2003) and
Whalen et al. (2003).  For this,
we implement an ``ionisation exclusion effect'' which will act to
inhibit sources from turning on before a certain amount of time has
passed since their gas was ionised. We express the recuperation time
$t_{\rm recup}$ in terms of the free fall time associated with each
source cell. In a spherical potential, the free fall time can be
expressed as,
\begin{equation}
t_{\rm ff} = \sqrt{\frac{3\pi}{32G\rho}},
\end{equation}
where $G$ is the gravitational constant and $\rho$ is the density.  We
note that free fall times are calculated only for the overdense cells
which actually contain sources, otherwise this time scale is not
meaningful.  As a first order approximation, we use this expression to
compute free fall times in source cells according to their matter
densities.  We can then express the recuperation time in terms of the
free fall time $t_{\rm recup} = K_{\rm recup} t_{\rm ff}$.  Here the
free parameter $K_{\rm recup}$ can be thought of as the relative
fraction of a free fall time that must pass before a source halo can
recuperate from the negative effects of photo-evaporation.  This
method requires us to carefully track when source cells become ionised
(either by their own radiation or by the radiation of neighbouring
sources) and subsequently increment the amount of time that passes in
these cells at each time step.  It should be noted that we also track
the migration of sources into different source cells during the course
of a particular exclusion interval so as to properly follow the total
wait time associated with each source halo.  Owing to the large
overdensities in source cells, recombination times are short and, as a
result, it is possible for a given source cell to undergo multiple
ionisations and inhibited phases. On a general level, this effect is
very similar to the ``volume exclusion effect'' adopted by Yoshida et
al. (2003b) although our method has the convenient feature of
naturally gauging the resiliency of each halo to be able to recuperate
from photo-heating. To explore both a weak and strong case of this
effect, we consider recuperation parameters $K_{\rm recup} = 1/3$ and
$K_{\rm recup} = 2/3$.

Next, we turn to photo-dissociation of molecular hydrogen by the LW
radiation from the first stars.  While photons with energies above
13.6 eV are likely to be completely absorbed by the gas surrounding
the sources, those with energies in the LW bands can easily travel
unimpeded through the neutral IGM and build up a uniform UV background
radiation field (Dekel \& Rees 1987; Haiman, et al. 2000; Omukai \&
Nishi 1999).  We emphasise that the net effect of this radiation is
{\it not} to completely inhibit primordial gas cooling, but it raises
the minimum mass scale of the halos in which the gas can cool (Haiman
et al. 2000; Machacek et al. 2001).  Recently, Yoshida et al. (2003c)
studied the effect of LW backgrounds on the formation of primordial
gas clouds and found that owing to the photo-dissociation of H$_2$,
gas cooling is suppressed for radiation with intensity $J_{21}>0.01$
(in units of $10^{21}$ erg s$^{-1}$ cm$^{-2}$ Hz$^{-1}$
str$^{-1}$). They also implemented a technique to compute H$_2$ column
densities around virialised regions making it possible to estimate gas
self-shielding factors in their maximal limits. With the help of
self-shielding, primordial gas in large halos can cool efficiently
even when exposed to background radiation with intensity $J_{21}
=0.01$. While the two extreme cases studied by Yoshida et al.
(2003c), an optically thin limit and maximal gas self-shielding,
should bracket the true effect of the LW background, a more careful
study is required to obtain an accurate estimate of the impact of this
radiation.

Given the uncertainties, we continue with our approach of adopting a
simple parameterisation aimed at capturing the qualitative aspects of
the effect. In particular, we first note that LW radiation is only
capable of affecting the abundance of molecular hydrogen in gas and
does not significantly affect the thermal state of the gas. Assuming
equilibrium, the amount of molecular hydrogen in a gas cloud is
inversely proportional to the photo-dissociation reaction coefficient
given by,
\begin{equation}
k_{\rm diss}=1.38\times 10^9 J_{\rm LW} K_{\rm shield},
\end{equation}
where $J_{\rm LW}$ is the radiation intensity at $h\nu=12.87$ eV and
$K_{\rm shield}$ is the effective shielding factor. Neglecting the
evolution of the gas density and temperature, and noting that the
cooling rate scales just as the number of hydrogen molecules, the
amount of gas that can cool should be proportional to
$1/(J_{\rm LW}K_{\rm shield})$.  Given this linear scaling we introduce a
modulating function,
\begin{equation}
F_{\rm LW} = a(J_{\rm LW}K_{\rm shield} - x_o) + b,
\end{equation}
from which the amount of cold gas available for star mass production
is computed after the effects of photo-dissociation are taken into
account.  To determine the fitting parameters $a$, $b$, and $x_o$, we
refer to Yoshida et al. (2003c) who point out that LW radiation is
unimportant for $J_{\rm LW} < 10^{-24}$, whereas for $J_{\rm LW} >
10^{-21}$, the gas cannot cool by molecular hydrogen cooling in nearly
all the halos in the optically thin limit ($K_{\rm
shielding}=1$). Thus by setting $a=10^{21}$, $b=1$, $x_o=10^{-24}$,
and restricting the maximum value of $F_{\rm LW}$ to 1, we are able to
mimic this behaviour.  The fact that the relevant range of intensities
is small lends support for our use of a linear approximation.

Assuming that an effective screen owing to abundant neutral hydrogen
blocks photons in the Lyman-series lines from all sources at redshifts
above $z_{\rm max}$ (Haiman et al. 1997), we can approximate $J_{\rm
LW}$ at each redshift $z_i$ by summing the emissivity in the LW band
from all previously active sources up to $z_{\rm max}$. For our
analysis, we set $z_{\rm max} =z_{i-1}$ and approximate $J_{\rm
LW}(z_i)$ as arising from the unattenuated radiation field purely from
the set of sources in the previous time step. Our assumption should be
reasonable given the large amount of neutral hydrogen present around
the time when $J_{\rm LW}$ becomes appreciable and the fact that
photons with $E=13.6$ eV will be redshifted out of the relevant energy
range (11.18-13.6eV) after $\sim18\%$ change in redshift. Since the
luminosity in the LW band per unit stellar mass for very massive stars
is $L_{\rm LW}\simeq 3\times10^{21}$ erg s$^{-1}$ Hz$^{-1}$
M$_{\odot}^{-1}$, with only a weak dependence on the stellar mass, our
expression for $J_{\rm LW}(z_i)$ becomes,
\begin{equation}
J_{\rm LW}(z_i) = \frac{c}{4\pi}~L_{\rm LW} \times
M_{\star}^{tot}(z_{i-1})\times \Delta t_i \times V_{\rm box}^{-1}(z_i),
\end{equation}
where $c$ is the speed of light, $M_{\star}^{tot}(z_{i-1})$ is the
total star mass which was switched on at redshift step $z_{i-1}$,
$\Delta t_i$ is our time step of $3\times 10^{6}$ yr, and
$V_{box}^{-1} (z_i)$ is the proper volume of our simulation box at
$z_i$.

Owing to the highly uncertain nature of shielding processes, we refrain
from developing a complex parameterisation of this effect (see,
however, \S 7.2 of Yoshida et al. 2003c). Rather, we adopt the
conservative value of $K_{\rm shield}=0.10$ as our maximum shielding
effectiveness for halos with $M_{\rm SF} > 10^5$ M$_\odot$ and linearly
scale this value with mass such that $K_{\rm shield}=1$ for $M_{\rm SF}\leq
10^3$ M$_{\odot}$ (corresponding to zero shielding). We apply this
level of shielding in all our models.

Finally, we turn to the effect of metal pollution from the life-cycles
of the first stars. Metals can cool gas
inside halos to temperatures lower than can be achieved by H$_2$
cooling, making possible the formation of less massive {\it ordinary}
population II stars. The critical transition between the two
populations should occur after a certain level of metal enrichment has
occurred. Assuming that the metallicity in the IGM is directly linked
to the amount of gas that has been incorporated into population III
stars, one can parameterise when this transition should occur in terms
of the fraction of total gas that forms massive stars.  Oh et
al. (2001) estimate that the critical transition should occur when a
fraction of $3\times 10^{-5} - 1.2\times 10^{-4}$ is formed into
massive stars in the range 150-250 M$_{\odot}$, which are believed to
be the main contributors of metals.  It should be noted, however, that
this estimate assumes that metals produced by massive stars are
uniformly mixed throughout the IGM, which is unlikely. If this is not
the case, and metals remain preferentially near sites of halo
formation, then we would expect that the critical fraction is
potentially much lower than the value reported by Oh et al. (2001). In
addition, stars with masses larger than 250 M$_{\odot}$ were excluded
under the assumption that their contribution to metal enrichment
should be small if they end their lives as black holes. However, the
amount of mass these stars shed before they collapse into black holes
is still relatively uncertain and may also reduce the fraction quoted
above.

Given the large uncertainty associated with this fraction, we allow
all our models to continually form population III stars all the way
to $z=16$ where we stop our simulations owing to the
growing non-linearity of the fundamental fluctuation mode.  For all
the models which we consider, the fraction of total gas mass that is
converted to stars $F_{\rm conv}$ by $z=16$ falls short of the lower bound
of the critical range quoted by Oh et al. (2001). In an effort to
study more protracted population III epochs, we will employ simple
extrapolations of the model results to lower redshifts. This will
allow us to assess the impact of population III stars for larger
$F_{\rm conv}$ values more consistent with the critical range quoted in Oh
et al. (2001) (see \S 4.3).

\subsection{Models}

Our goal of identifying the important features which are associated
with the ability of population III stars to reionise the Universe
leaves us with a rather large parameter space to explore. Namely, even
neglecting uncertainties associated with the LW background and metal
enrichment, we are still left with 4 free parameters: $f_{\rm esc}$,
$E_{\rm SF}$, $D_{\rm f}$, and $K_{\rm recup}$. Our approach of
considering both a weak and strong case for each of these parameters
thus leads to 16 separate models which we list in Table 1.

In addition, we include a further model M9 which will represent our
most restrictive case where we allow only a single 200 M$_{\odot}$
star to form in each halo which has at least $4\times 10^4$
M$_{\odot}$ of star forming mass ($E_{\rm SF}=0.005$). This model is
meant to be more consistent with the picture described by Abel et
al. (2002) who have used high resolution hydrodynamical simulations to
show that a pre-galactic halo tends to form only a single collapsing
core in its center without renewed fragmentation. In an effort to
simulate the ultimate inhibition of subsequent star formation in these
halos, we further restrict the model to allow only one episode of
population III star-formation per halo without the possibility of
forming new metal-free in the same halo or its descendents at a later
time. In this case, once a single source turns on within the halo or
its gas becomes ionised by a neighbouring source, we inhibit it from
hosting any future metal-free stars ($D_{\rm f} =\infty$, $K_{\rm
recup}=\infty$). It is important to point out that the total number
of ionising photons released by a single 200 M$_{\odot}$ star in the
course of $3\times10^6$ yrs easily exceeds the amount of baryons in
even our most massive halos and therefore justifies our use of
relatively large escape fractions in single-star halos (see Whalen et
al. 2003).

The main goal of our analysis will be to explore the effectiveness of
population III stars as ionising sources in the early Universe.  By
considering a plausible set of models which incorporate feedback
effects to varying degrees, we develop a better qualitative
understanding of the complex interplay between these processes and
reionisation. Our inclusion of an {\it ultra-restrictive} model will
allow us to loosely constrain the minimum effectiveness of population
III reionisation.

\section{RESULTS AND DISCUSSION}

The reionisation history of H {\small I} and He {\small II} is
followed in each of our models using the radiative transfer code
described in Paper I.  Briefly, radiative transfer is
performed on a Cartesian grid with $200^3$ cells using an adaptive
ray-casting scheme (Abel \& Wandelt 2002). Density fields, clumping
factors, and source characteristics are taken from outputs of the
hydrodynamic simulations; thus we neglect the dynamical feedback of
the radiation on structure formation\footnote{See Sokasian, Abel \&
Hernquist (2001) for a more detailed description of how we integrate
radiative transfer calculations with existing outputs from
cosmological simulations.}. However, to account for the neglected
photo-heating of the IGM, we compute recombinations in our radiative
transfer calculations according to an artificially raised temperature
of $T=1.5\times10^4$ K. This temperature corresponds to a reasonable
estimate of the IGM temperature after reionisation by sources with
spectra typical of metal-free sources (Hui \& Haiman 2003). If the
photoionised gas has a higher temperature, recombinations occur at a
slightly slower rate (the recombination time $t_{\rm rec}\propto T^{0.7}$)
and ionisation zones grow marginally faster. For lower temperatures,
the opposite is true.

In the context of reionisation studies, cosmic variance limits our
ability to capture both a representative mass function for source
halos and also a reliable picture of the distribution of gas in the
Universe. Since our simulation box has a comoving length of only 1
Mpc, we will not be able to make very strong statements regarding the
precise moment of cosmic reionisation. However, as stated earlier, our
goal is only to assess the relative properties of a set of physically
motivated models for the impact population III sources can have on
reionisation. Having stated this, we note that the same small box
simulation was analysed in Yoshida et al. (2003a) and was found to
contain halo abundances for $M>7\times10^5$ M$_{\odot}$ that were in
reasonable agreement with the Press-Schechter mass function between
$17<z<25$ (see also Jang-Condell \& Hernquist 2001).  
The incomplete sampling of the halo mass function due to
the finite box size is appreciable only at $z>30$. However, as we
shall show, star formation in our models is not dominated by rare
massive halos and therefore simulating larger volumes should not
significantly alter our results.

\subsection{Star Formation History}
In Figure 3, we show the evolution of the star formation rate (SFR)
density for population III stars for all our models. Here the SFR is
computed on the fly as the simulation regulates the amount of star
mass that actually turns on after incorporating the effects of
feedback. By $z=16$, all but our most restrictive model M9 appear to
reach a SFR near $10^{-3}$ M$_{\odot}$ yr$^{-1}$ Mpc$^{-3}$. The
similarity in the SFRs between the different models brings attention
to the self-regulating consequences of feedback.  For example, when
the disruption factor is large, less star formation takes place
locally, which in turn minimises negative feedback in the form of
dissociating radiation and photo-evaporation. As a result, star
formation is promoted externally and the global SFR is regulated. In a
similar fashion, models with a low escape fraction limit the effect of
photo-evaporation externally and therefore are able to have a similar
level of star formation as models with a large escape fraction.  In \S
4.3 we shall take a more detailed look at dynamic feedback and attempt
to quantify the effect.

Nevertheless, it is important to point out that a SFR of $10^{-3}$
M$_{\odot}$ yr$^{-1}$ Mpc$^{-3}$ is roughly an order of magnitude
lower than the ``normal-mode'' (population II) star formation
predictions from Springel \& Hernquist (2003) and Hernquist \&
Springel (2003) at the same redshift. Here, ``normal mode''
refers to star formation occurring in larger mass systems
where gas cooling takes place via atomic hydrogen and helium
transitions and the only form of regulating feedback is
supernovae. Besides the difference in amplitude, the SFR in our
population III models appears to have relatively ``flatter'' shapes in
the range $16<z<20$ compared to the ``normal-mode'' SFR of Hernquist
\& Springel (2003).  This is not surprising given the stronger
regulating processes associated with population III star formation.
The overall shape and the amplitude of the SFR are in reasonable
agreement with the semi-analytic prediction of Yoshida et al.  (2003c,
their Fig. 18).  It is important to note that population III star
formation at $z\lesssim20$ may have a significant impact on subsequent
``normal-mode'' star formation occurring in small proto-galaxies which
are susceptible to the negative feedback effects of photoheating.
Because these effects were not included, it is not surprising that the
population III SFR presented here does not converge with the
``normal-mode'' SFR of Hernquist \& Springel (2003).  Nevertheless, if
the epoch of population III stars ends abruptly, as we have assumed
here, then the transition itself should also be abrupt. A very
interesting issue for future work, therefore, would be to address how
the transition from population III stars to ordinary star formation
occurs. The discontinuity presented here represents an additional
uncertainty of our analysis only when we attempt to compute a complete
history of reionisation which spans both population II and population
III stars.

\subsection{Global Ionisation Fractions}

The evolution of the global ionisation fraction in the simulation
volume is a useful tool for characterising differences between the
various models. In Figures 4 and 5 we show the evolution of the
volume-weighted ionisation fraction down to $z=16$ for H {\small II}
and He {\small III}, respectively. In each panel, we plot both the
$f_{\rm esc}=1.0$ ({\it solid-line}) and $f_{\rm esc}=0.3$ ({\it
dashed-line}) cases for the corresponding model. Here, we note that
with the exception of our most restrictive model M9, all the $f_{\rm
esc}=1.0$ models are able to significantly ionise their volumes by
$z=16$.  Interestingly, some of the models with higher emissivity are
also able to substantially doubly ionise helium although the
corresponding fractions are much less than in H {\small II}. This is
true despite the fact that the ionising spectrum of a typical $300$
M$_{\odot}$ metal-free star produces roughly $\sim 3$ times more
ionising photons per helium atom compared to hydrogen for cosmic
abundances. The difference in the resultant ionisation fractions
owes to the fact that the recombination rate of He {\small III} is much
larger relative to H {\small II}. This large recombination rate is
also responsible for making He {\small III} ionisation evolution
relatively less smooth, because relic He {\small III} regions are much
quicker to recombine.

In all our models, the initial rise of the ionisation fraction appears
to be closely correlated with the exponential growth of star-forming
halos in the volume. Interestingly, however, our highest emissivity models
(M1$_{\rm b}$, M2$_{\rm b}$, M3$_{\rm b}$) appear to show some degree of
divergence from this rate at later times. In particular, the rise of
the ionisation fraction becomes shallower once the ionisation fraction
exceeds $\sim 60\%$. The mostly likely explanation for this behaviour
is that dynamic external feedback becomes particularly inhibitive once
a substantial portion of the volume has become ionised and many
sources become involved in the process.  We will explore this effect
in more detail in the next section where we try to assess the impact
from this dynamic component of the feedback in terms of the amount of
star mass that actually turns on at each time step.

A summary of the ionisation results at $z=16$ is given in Table 2
which lists for each model: the fraction of total gas mass converted
to stars $F_{\rm conv}$, the cumulative number of ionising photons
released per atom, and the volume ionisation fraction for H {\small
II} and He {\small III}, respectively. Here we point out that while
our more restrictive models do not achieve as high of an ionisation
fraction as the models with larger emissivities by $z=16$, they
convert less gas mass into stars and therefore should have polluted
their environment relatively less with metals. As a result, we would
expect that metal-free star formation should persist longer in these
models.  It is particularly interesting to compare the results for our
most restrictive model M9 to our most liberal model M1. We note in
particular that the realisation of a ``one-star-per-halo'' for
metal-free stars as defined in our M9 model has such a slow rate of
growth for its ionising emissivity that recombinations will likely
prevent the ionisation fraction from increasing significantly even if
we had extended this model down to lower redshifts. However, we point
out that this model is relatively more efficient in terms of the
resultant ionisation fraction per ionising photon than the M1 model
with larger emissivity.  This feature can be attributed to the fact
that models with higher emissivity suffer cumulatively more
recombinations as a result of ionising substantial portions of their
volumes at very early redshifts and keeping them ionised down to lower
redshifts. It should be noted, however, that this difference will be
slightly diminished by the fact that ionisation fronts in the more
restrictive M9 model have difficulty breaking through the dense
environments surrounding the sources. This causes the M9 model to
suffer relatively more recombinations {\it per unit volume} than
models with higher emissivity whose ionisation zones expand into the
less dense and more voluminous portions of the IGM. This morphological
effect also helps to explain why models M4-M9, which have not yet
experienced a slow-down in the rate of ionising photon production (owing
to external feedback factors), exhibit ionisation fractions for their
$f_{\rm esc}=0.3$ case that are systematically lower than $0.3$ times
the ionisation fractions predicted in the corresponding $f_{\rm
esc}=1.0$ case (see Table 2).

A visual illustration of the reionisation process is shown in Figure 6
where we plot a series of projected slices through the simulation
volume from the M1$_a$ model (other models exhibit a similar
topological evolution in their ionisation structures).  In each panel,
a $0.25$ Mpc slice is projected in both density and
ionisation fraction.  From the plots one can follow the growth of the
ionisation zones ({\it blue}) around the first stars as they turn
neutral gas ({\it yellow}) into highly ionised regions ({\it
light-blue}). The plots clearly show the preferential advance of
ionisation fronts into the underdense regions, responsible for the
steep rise seen in the corresponding volume-weighted ionisation
fractions.

To understand which halos are contributing the bulk of the ionising
radiation, we plot in Figure 7 the fraction of ionising flux released
as function of halo mass for three redshift ranges associated with the
M1$_{b}$ run (other models had similar results). Here we see that
through the redshift ranges $21\leq z<23$, $19\leq z<21$, and $16\leq
z<19$, the bulk of ionising radiation is consistently released from
halos with masses between $\sim 2\times 10^{6}$ M$_{\odot}$ and $\sim
6\times 10^{6}$ M$_{\odot}$. The fact that these ``mini-halos'' are
capable of substantially ionising the IGM appears to contrast with the
findings of Ricotti et al. (2002a) who conducted similar 3D
cosmological simulations which included dynamically linked radiative
transfer calculations (see Ricotti et al. 2002b for simulation
details). They showed that negative feedback prevents the size of H
{\small II} regions from exceeding the size of the dense filaments;
hence, ``mini-halos'' are not able to reionise the voids. More
specifically, the Ricotti et al. (2002a) simulations showed that
positive feedback from the accelerated formation rate of molecular
hydrogen in front of ionisation fronts within these filaments provided
an effective shield preventing the destruction of molecular hydrogen
in other star forming regions. On the contrary, when the ionisation
fronts expand beyond the filaments, densities drop and the shield
disappears allowing negative feedback from LW radiation and
photoevaporation to dominate. Ricotti et al. (2002a) thus concluded
that the volume filling factor of the H {\small II} regions associated
with these ``mini-halos'' remains small and cannot reionise the
universe.

Although none of the models considered in our analysis are able to fully
reionise the universe before the onset of population II stars, they do
ionise a substantial portion of the IGM. This discrepancy most likely
results from the way in which negative feedback is treated in the
simulations. More specifically, by dynamically linking approximate
radiative transfer calculations directly with cosmological simulations,
Ricotti et al. (2002a) are able to simulate positive feedback processes
preceding H {\small II} regions which effectively reduces negative feedback
from the dissociation of molecular hydrogen. However, when the ionising
emissivity of ``mini-halos'' is large, ionisation fronts expand outside of
the filaments and negative feedback from dissociation and photo-heating
begins to dominate. Since our analysis does not allow us to couple
radiative transfer effects directly into the cosmological simulations, we
are unable to include positive feedback effects such as the one mentioned
above. Nevertheless, the models we consider are able to easily force
ionisation fronts beyond the dense filaments and therefore we do not expect
positive feedback to be important in these cases. Once ionisation fronts
expand into the IGM, negative feedback effects kick in through our {\it
  ad-hoc} implementation of the photo-evaporation effect and the LW
dissociating background. As we shall show in the following section, such
feedback works to regulate further star formation, however, unlike Ricotti
et al. (2002a) we do not find that it fully suppresses star formation. It
is important to note that our simulations included an approximate treatment
which allowed halos to shield themselves against dissociating radiation in
proportion to their masses.  In contrast, self-shielding was excluded in
the simulations of Ricotti et al. (2002a) and may have resulted in an
overestimation of negative feedback, plausibly causing much of our
discrepancy with the results by Ricotti et al. (2002a). Furthermore, it
should be pointed out that the discrepancy may also be related to a number
of other differences between the two simulations such as the differing
methodologies used in the calculation of the star forming mass in the halos
as well as differences in the choice of the IMF.

\subsection{Dynamical Feedback from the LW Background and 
        Photo-evaporation} 

In Figures 4 and 5 we showed the evolution of the global
volume-weighted ionisation fraction for each model.  While useful for
making comparisons between the net ionising ability of the models, the
figures do not reveal the complex role dynamical feedback may have
played in the evolution. In this section we attempt to study the
effectiveness and subsequent evolution of this feedback more
directly. In particular, we focus on tracking the modulating behaviour
of our implementation of the LW background and the photo-evaporation
effect (or in terms of our parameterisation, the ``ionisation
exclusion effect'').  More specifically, by tracking the ratio of the
star mass that actually turns on at each time step $M_{\rm star}^{\rm
on}$ to the star mass originally available $M_{\rm star}^{\rm orig}$
(computed in the absence on any external feedback), we can directly
study the net impact from both these effects. In Figure 8 we show the
ratio $M_{\rm star}^{\rm on}/M_{\rm star}^{\rm orig}$ at each redshift
step as a histogram for all our models. The plots reveal sharp swings
mainly owing to the discrete timesteps used. However, these swings are
enhanced at early times, reflecting the strong nature of our
``ionisation exclusion effect'' when localised to a small region with
only a few sources, which is the case when the first cluster of
sources turns on. As more sources are invoked at lower redshifts, this
ratio begins to behave more smoothly.

In Table 3 we list for all the models the mean value of $M_{\rm
star}^{\rm on}/M_{\rm star}^{\rm orig}$ weighted by the original star
mass between the time the first source turns on and $z=16$.  This
mass-weighted mean quantifies the relative effectiveness of external
dynamical feedback. Note, in particular, how models in which we have
incorporated strong local feedback effects have correspondingly low
values for this mean ratio (compare, for example, models
M1$\leftrightarrow$M2 and M3$\leftrightarrow$M4). This is a clear
indication of the self-regulating feature of the population III era:
when local negative feedback is strong, the global component of
negative feedback is reduced, allowing for more sources to turn on in
distant regions. This feature helps to explain how such a large set of
model realisations leads to the relatively similar ionisation
histories seen in Figures 4 and 5.

Bearing in mind that the net effect of the LW background is to
systematically reduce $M_{\rm star}^{\rm on}/M_{\rm star}^{\rm orig}$
in proportion to the star mass that froms, one can use the mean of
this ratio to isolate and assess the relative impact of the
photo-evaporation effect between the models M1$\leftrightarrow$M5,
M2$\leftrightarrow$M6, M3$\leftrightarrow$M7, and
M4$\leftrightarrow$M8. Note in particular, the significant drop in
this ratio from M1$\rightarrow$M5, which gives an indication that
external feedback may play an important role in scenarios where star
formation is efficient and internal feedback is low.

Interestingly, the effectiveness of external feedback is strong by
$z\sim16$ even in model M9 which turns on very few sources.  This is
most likely owing to the fact that these sources reside in only the most
clustered regions containing massive enough halos capable of hosting
sources in this {\it ultra-restrictive} model. The close proximity of
the sources to one another in these regions therefore keeps external
feedback in the form of photo-evaporation particularly effective in
inhibiting a fair portion of the sources from turning on.
We mention that, under certain conditions, weak ionising radiation
can promote molecular hydrogen formation and 
enhance primordial gas cooling, hence causing possibly 
{\it positive} feedback effects (Haiman, Rees, \& Loeb 1996; 
Kitayama et al. 2001). We do not consider the effects in our models
and thus all the models may be restrictive in this sense.

\subsection{Electron Optical Depth: Population II and III sources}

The recent tentative measurement of polarisation in the CMB by the
WMAP satellite (Kogut et al. 2003) indicates that
reionisation occurred earlier than implied by the SDSS quasars alone
(e.g., Spergel et al. 2003).  The optical depth to Thomson scattering
is $0.13 < \tau_e < 0.21$ corresponding to an instantaneous
reionisation epoch between $14<z<20$.  In Paper I, we showed that
stellar sources similar to those seen in local galaxies;
i.e. population II type stars, are unable to reionise the Universe by
such high redshifts. Meanwhile, the analysis in this paper indicates
that metal-free (population III) stars could have significantly
reionised the Universe at $z>16$.  Cen (2003a,b) and Wyithe \& Loeb
(2003a,b) explored a large number of models of reionisation history
using semi-analytic methods and argue that, in some specific cases,
the Universe could be reionised twice.

In an effort to study this scenario more quantitatively, we would like
to couple the ionisation histories from both types of sources and
compute the resulting integrated electron optical
depth. Unfortunately, owing to the high resolution necessary for
carrying out simulations involving population III stars, we are unable
to directly include these sources in the larger simulation volumes
necessary for studying cosmic reionisation by population II sources
(galaxies). We are therefore forced to adopt the more approximate
approach of linking together the ionisation histories from the two
sets of simulation volumes.  More specifically, our approach relies on
using the results of our population III simulations to estimate the
volume-weighted ionisation fraction at redshifts around which it is
reasonable to assume that star-forming halos have exhausted their
ability to form additional metal-free stars.  At this point in time,
we then re-run the $f_{\rm esc}=0.20$ population II simulation from
Paper I (conducted in a 14.3 Mpc comoving box) with the initial
conditions that the IGM was {\it uniformly} ionised to the same
ionisation fraction (both in hydrogen and helium) as estimated by our
population III simulations and also {\it uniformly} heated to
$T=1.5\times10^4$ K. Our use of this particular model for population
II sources is motivated by our analysis in Paper I where we found that
given the amplitude and form of the underlying star formation
predictions, an escape fraction near 20\% is most successful in terms
of statistically matching observational results from the $z=6.28$ SDSS
quasar of Becker et al. (2001) (see Paper I for further details). By
linking together the two simulations in this fashion, we are assuming
that the onset of population II begins exclusively at redshifts below
the transition redshift and that there are no population III stars
thereafter.  Since the relative contribution predicted for population
II sources in the simulations is small above the transition redshifts
which we will consider, we can reliably separate the two contributions
in this manner.

Given the fact that most of our population III models
ionise the volume to a similar extent by $z=16$, we are motivated to
simplify our analysis by grouping together these models into a single
model by averaging all the results at each redshift.  We perform this
averaging for models M1-M8, but continue to keep model M9 distinct as
it represents our {\it ultra-restrictive} case which predicts
significantly lower ionisation fractions as a result of invoking only
a single 200 M$_{\odot}$ star once per star forming halo. By $z=16$,
at which point we stop our population III simulations,
our combined model M1-M8
with $f_{\rm esc}=0.30$ ($f_{\rm esc}=1$) has converted a fraction
$F_{\rm conv}=1.53\times 10^{-5}$ ($F_{\rm conv}=1.26\times 10^{-5}$) of its
total gas into stars.  This fraction falls short of the lower limit
estimated by Oh et al. (2001) for the transition to the population II
epoch ($\sim 3\times 10^{-5}$). In an effort to examine a more
prolonged population III epoch culminating with gas-to-star conversion
fractions more in-line with the estimates from Oh et al. (2001), we
apply a simple extrapolation of all the simulation results to
redshifts below $z=16$. More specifically, we extrapolate the simulations
using the associated rate of change in the models between
$z=18$ and $z=16$. For the combined M1-M8 model with $f_{\rm esc}=0.30$
($f_{\rm esc}=1$), we extrapolate the population III epoch down to
$z=14.5$ and $z=12.0$ resulting in gas conversion fractions:
$F_{\rm conv}=2.20\times 10^{-5}$ ($F_{\rm conv}=1.90\times 10^{-5}$) and
$F_{\rm conv}=3.32\times 10^{-5}$ ($F_{\rm conv}=2.86\times 10^{-5}$),
respectively. In the case of model M9 with $f_{\rm esc}=0.30$
($f_{\rm esc}=1$), even an extrapolation down to $z=12$ only yields a
conversion fraction of $F_{\rm conv}=5.10\times 10^{-6}$
($F_{\rm conv}=3.94\times 10^{-6}$), well below the estimate from Oh et
al. (2001).  However at $z\lesssim12$, population II stars which form
via atomic line cooling in massive halos begin to dominate the
ionising emissivity relative to the population III stars incorporated
in M9. We therefore adopt $z=12$ as our epoch transition redshift for
this model keeping in mind that the exclusion of population III
sources below $z=12$ slightly underestimates the resultant rise in the
global ionisation fraction.

In Figure 9 and 10, we plot the evolution of the H {\small II}
volume-weighted ionisation fraction ({\it top-panel}) and the
corresponding optical depth to Thomson scattering $\tau_e$ between the
present and redshift $z$ ({\it bottom-panel}) for the models discussed
above with $f_{\rm esc}=0.3$ and $f_{\rm esc}=1.0$, respectively.
Here the vertical dashed-dotted line represents the redshift ($z=16$)
where we stop our population III simulations.  The meaning of the
various curves is summarised in the caption to Figure 9.  The electron
optical depth was calculated from
\begin{equation}
\tau_e(z)=\int^z_0\sigma_T~ n_e(z^{\prime})~c ~\left|
\frac{dt}{dz^{\prime}} \right| ~dz^{\prime},
\end{equation}
where $\sigma_T=6.65\times10^{-25}$ cm$^2$ is the Thomson cross
section and $n_e(z^{\prime})$ is the mean electron number density at
$z^{\prime}$. In the calculation of $\tau_e$ we use mass-weighted
ionisation fractions and assume that helium is singly ionised to the
same degree as the hydrogen component for $z>3$ and doubly ionised
everywhere at $z<3$. The figure clearly demonstrates how population
III sources may have significantly contributed to the electron optical
depth, especially in the case where the escape fraction from the
mini-halos hosting metal-free stars is close to unity. Interestingly,
the relatively late onset of population III stars does not cause a
dramatic recombination era before population II stars begin to
continue the reionisation process.

Comparisons with the WMAP results show that only the combined model
with $f_{\rm esc}=1.0$ comes close to the inferred mean value of
$\tau_e=0.17$ ({\it dashed-line}) with optical depths $0.141$ and
$0.152$ for population III extrapolations to $z=14.5$ and $z=12$,
respectively. In the case of our {\it ultra-conservative} M9 model
with $f_{\rm esc}=1.0$, the resultant optical depth is $\sim40\%$ larger
than in the case with only population II stars only, but is slightly
shy of the lower $2\sigma$ limit of the WMAP measurement. In all cases,
however, it appears that a large escape fraction is required during
the population III epoch in order to significantly contribute to the
integrated optical depth.
 
\section{SUMMARY AND CONCLUSIONS}

In this paper we have applied approximate radiative transfer
calculations to the results of cosmological simulations capable of
following early structure formation in an effort to study the
potential impact that the first generation of metal-free stars may have had
on the reionisation history of the Universe. By incorporating a series
of approximate feedback effects, we are able to simulate the complex
interrelated processes which may have self-regulated subsequent
population III star formation. We have examined a set of models in
which the severity of the effects was varied in order to capture
qualitatively how the feedback can affect star-formation at early
epochs. 

Overall, we find that for a plausible range of values related to our
parameterisations, population III star formation proceeds in a
self-regulated manner. As a result, the net impact of these sources in
terms of reionisation is fairly insensitive to the relative degree of
severity of one form of feedback over another. This is true as long
internal feedback within a halo is not exceedingly large and halos
can host multiple generations of population III stars. In particular,
we find that if we restrict halos of sufficient mass
to host only a single $200$ M$_{\odot}$ star {\it once} in their
existence, then the impact of population III sources on reionisation
is significantly reduced.

In the models where there is no restriction on the number of stars a
halo can host and the escape fraction is unity, the mean level of
ionisation reached by $z=16$ is $66\%$ and $29\%$ for H {\small II} and
He {\small III}, respectively.  One would therefore expect that at
least within the context of this description, population III stars
should have had a significant impact on the ionisation history of the
Universe. Interestingly, the fact that the helium component may also
have been ionised to a significant fraction has
consequences for the thermal history of the IGM (e.g. Hui
\& Haiman 2003) and could possibly be observable with future
observations of He {\small II} recombination lines in the host halos
using the Next Generation Space Telescope (Oh, Haiman, \& Rees 2001; 
Tumlinson, Giroux, \& Shull 2001).

By linking together the simulation results of the population II
study from Paper I with the results of the population III
analysis presented here, we are able to synthesise a full history of
how the Universe may have been reionised.  We find that even with
fairly conservative estimates for negative feedback, population III
sources with large escape fractions are able to significantly raise
the total electron optical depth so that it becomes statistically
consistent (within the 1$\sigma$ lower bound) with the WMAP
measurement.  
We note that the Thomson optical depths predicted in our
analysis appear to be lower than those obtained in the semi-analytic
treatment conducted by Cen (2003a). Correspondingly, population III
sources in our analysis appear to significantly affect the
reionisation history of the Universe only at relatively later times
($z\lesssim19$) than the models considered by Cen (2003a). Nevertheless,
both studies appear to suggest that achieving $\tau_e>0.17$ seems
quite difficult with reasonable assumptions for the properties of
population III stars. It is impractical to make more direct comparisons
between the two works given the very different approaches employed and
the large number of parameters. It should also be noted that
semi-analytic treatments do no allow for the direct incorporation of
negative feedback effects and instead absorb these uncertainties
directly into star formation efficiency factors. As a result, they may
oversimplify the true impact of these effects if the amount of star
mass that turns on in each halo is a sensitive function of halo
properties such as proximity to other halos, merger history, and total
mass (Yoshida et al. 2003c).

In carrying out our analysis, we were forced to extrapolate a small
portion of the population III ionisation history to lower redshifts
owing to boxsize limitations. In doing so, we were able to produce
gas-to-star conversion fractions which approached the critical range
described in Oh et al. (2001), estimated to mark the transition to the
population II epoch. It is important to point out that the simple
recipe of Oh et al. (2001) may not accurately reflect the transitional
stage if metal pollution is not efficient. In particular, in the case
of an inefficient mixing mechanism, the metallicity in overdense
regions can almost reach solar levels even at $z>16$ for a top heavy
IMF, preemptively ending the population III epoch (e.g. Yoshida et al.
2003e). 
Nevertheless, in all
of our models we switched over to a population II epoch at
aggressively conservative values below or slightly above the lower
bound of the estimated range in Oh et al. (2001). Therefore, it is
plausible that population III star formation continues beyond the
transitional redshifts considered here. If that is the case,
population III stars would continue to significantly ionise the IGM,
potentially leading to a single reionisation epoch in which population
II sources turn on at relatively late times and are intense enough to
keep the IGM in a highly ionised state. In fact, our combined $f_{\rm
esc}=1$ population III model, when extrapolated down to $z=12$,
exhibits such a single reionisation epoch. Interestingly, the results
from this model appear very similar to the prediction from our model
in Paper I which incorporated an evolving ionising boosting factor as
an additional multiplicative factor to the intrinsic ionisation rates
in population II sources. Such an evolution would be indicative of an
evolving IMF which became more and more top-heavy with increasing
redshift.  We note however, that such a ``vanilla'' model of
reionisation may be inconsistent with the observational constraints
from the thermal history of the IGM as noted earlier. Future
observations of CMB polarisation by the Planck surveyor, potentially
tractable observations of the 21 cm emission signal from neutral
hydrogen at high redshifts (see Furlanetto et al. 2003; Ciardi \&
Madau 2003; Zaldarriaga et al. 2003),
and searches for Ly-$\alpha$ emission from galaxies at the 
reionisation epoch
(e.g. Barton et al. 2003) may help resolve these degeneracies.

Through our analysis, we have attempted to assess the potential role
feedback effects may have had on the ability of population III sources
to reionise the Universe.  While our radiative transfer simulations
capture nearly all the important physical mechanisms operating during
reionisation, the precise effects of some processes that are modeled
in an {\it ad hoc} manner remain still uncertain.  More accurate
studies may soon be forthcoming with the incorporation of radiative
transfer calculations directly into cosmological simulations similar
to the one used here.

Finally, we emphasise that the results here were obtained for a
conventional $\Lambda$CDM cosmology.  If the true power spectrum is
actually similar to the running spectral index (RSI) model advocated
by Spergel et al. (2003), ``first star formation'' in mini-halos will
be greatly suppressed (Yoshida et al. (2003a).  It is an open question
whether the first luminous sources at redshifts below $z\simlt 18$ in
a RSI cosmology can produce a Thomson optical depth consistent with
the WMAP measurement.

\begin{acknowledgments}

We thank Volker Bromm and Marie Machacek for insightful discussions
related to this study.  This work was supported in part by NSF grants
ACI 96-19019, AST 98-02568, AST 99-00877, and AST 00-71019 and NASA
ATP grants NAG5-12140
and NAG5-13292.  NY acknowledges support from the Japan Society
of Promotion of Science Special Research Fellowship.  The simulations
were performed at the Center for Parallel Astrophysical Computing at
the Harvard-Smithsonian Center for Astrophysics.

\end{acknowledgments}

\clearpage

\begin{table}[htb]
\begin{center}
\begin{tabular} {lccccc}
\multicolumn{6}{c}{\textbf{TABLE 1 }} \\
\multicolumn{6}{c}{Models} \\
\hline
\hline
Model         &\vline & $f_{\rm esc}$ & $E_{\rm SF}$ & $D_f$ & $K_{\rm recup}$ \\
\hline
M1$_{(a,b)}$  &\vline& (0.3)$_a$ (1.0)$_b$  & 0.02   & 40  & 1/3 \\   
M2$_{(a,b)}$  &\vline& (0.3)$_a$ (1.0)$_b$  & 0.02   & 120 & 1/3 \\
M3$_{(a,b)}$  &\vline& (0.3)$_a$ (1.0)$_b$  & 0.005  & 40  & 1/3 \\
M4$_{(a,b)}$  &\vline& (0.3)$_a$ (1.0)$_b$  & 0.005  & 120 & 1/3 \\
M5$_{(a,b)}$  &\vline& (0.3)$_a$ (1.0)$_b$  & 0.02   & 40  & 2/3 \\
M6$_{(a,b)}$  &\vline& (0.3)$_a$ (1.0)$_b$  & 0.02   & 120 & 2/3 \\
M7$_{(a,b)}$  &\vline& (0.3)$_a$ (1.0)$_b$  & 0.005  & 40  & 2/3 \\
M8$_{(a,b)}$  &\vline& (0.3)$_a$ (1.0)$_b$  & 0.005  & 120 & 2/3 \\
M9$_{(a,b)}$  &\vline& (0.3)$_a$ (1.0)$_b$  & 0.005  & $\infty$ & $\infty$ \\
\hline
\end{tabular}
\end{center}
\end{table}

\clearpage
\begin{table}[htb]
\begin{center}
\begin{tabular} {lcccccc}
\multicolumn{7}{c}{\textbf{TABLE 2 }} \\
\multicolumn{7}{c}{Summary of model results at $z=16$} \\
\hline
\hline
        &\vline & &$\sum~${{\small H II} phtns}  &$\sum~${\small He III} phtns & {\small H II} & {\small He III}   \\
Model         &\vline & $F_{\rm conv}$ &  per {\small H} atom &  per He atom &$X_{\rm vol}$ & $X_{\rm vol}$  \\
\hline       
M1$_{(a,b)}$ &\vline & (2.3e-05)$_a$ (2.1e-05)$_b$ & (0.87) (2.64) & (2.16) (6.59) & (0.39) (0.88) & (0.13) (0.67) \\ 
M2$_{(a,b)}$ &\vline & (1.5e-05)$_a$ (1.4e-05)$_b$ & (0.56) (1.71) & (1.36) (4.18) & (0.19) (0.75) & (0.09) (0.36) \\ 
M3$_{(a,b)}$ &\vline & (2.2e-05)$_a$ (1.9e-05)$_b$ & (0.79) (2.39) & (1.95) (5.93) & (0.36) (0.85) & (0.11) (0.55) \\ 
M4$_{(a,b)}$ &\vline & (1.4e-05)$_a$ (1.2e-05)$_b$ & (0.44) (1.28) & (0.94) (2.82) & (0.14) (0.67) & (0.04) (0.20) \\ 
M5$_{(a,b)}$ &\vline & (1.4e-05)$_a$ (1.2e-05)$_b$ & (0.51) (1.50) & (1.25) (3.71) & (0.18) (0.71) & (0.08) (0.34) \\ 
M6$_{(a,b)}$ &\vline & (1.1e-05)$_a$ (9.6e-06)$_b$ & (0.41) (1.16) & (0.99) (2.82) & (0.13) (0.62) & (0.07) (0.28) \\ 
M7$_{(a,b)}$ &\vline & (1.3e-05)$_a$ (1.0e-05)$_b$ & (0.45) (1.23) & (1.08) (2.98) & (0.16) (0.67) & (0.07) (0.29) \\ 
M8$_{(a,b)}$ &\vline & (9.9e-06)$_a$ (7.8e-06)$_b$ & (0.32) (0.86) & (0.70) (1.94) & (0.10) (0.44) & (0.03) (0.16) \\ 
M9$_{(a,b)}$ &\vline & (2.4e-06)$_a$ (1.9e-06)$_b$ & (0.09) (0.24) & (0.22) (0.59) & (0.03) (0.12) & (0.03) (0.13) \\ 
\hline
\end{tabular}
\end{center}
\end{table}

\clearpage

\begin{table}[htb]
\begin{center}
\begin{tabular} {lcc}
\multicolumn{3}{c}{\textbf{TABLE 3 }} \\
\multicolumn{3}{c}{Mean fraction for dynamic modulation of star mass} \\
\hline
\hline
 
Model &\vline &
$\text{mean}[\frac{ M_{\rm star}^{\rm on}}{M_{\rm star}^{\rm orig}}]_{\text{mass-weighted}}$ \\
\hline
M1$_{(a,b)}$   &\vline& (0.46)$_a$ (0.44)$_b$ \\
M2$_{(a,b)}$   &\vline& (0.67)$_a$ (0.62)$_b$ \\
M3$_{(a,b)}$   &\vline& (0.54)$_a$ (0.48)$_b$ \\
M4$_{(a,b)}$   &\vline& (0.68)$_a$ (0.56)$_b$ \\
M5$_{(a,b)}$   &\vline& (0.28)$_a$ (0.24)$_b$ \\
M6$_{(a,b)}$   &\vline& (0.48)$_a$ (0.41)$_b$ \\
M7$_{(a,b)}$   &\vline& (0.33)$_a$ (0.28)$_b$ \\
M8$_{(a,b)}$   &\vline& (0.50)$_a$ (0.39)$_b$ \\
M9$_{(a,b)}$   &\vline& (0.59)$_a$ (0.41)$_b$ \\
\hline
\end{tabular}
\end{center}
\end{table}

\clearpage

\begin{figure}[htb]
\figurenum{1}
\setlength{\unitlength}{1in}
\begin{picture}(6,6.5)
\put(0.90,-1.8){\includegraphics{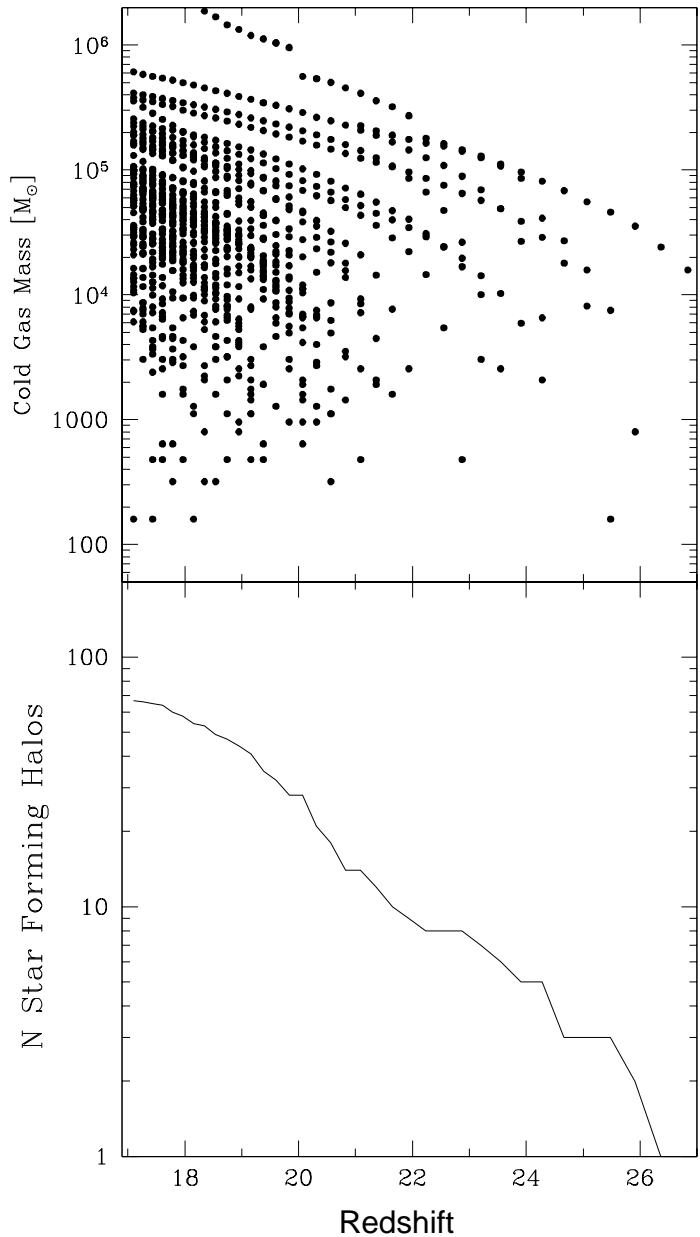}}
\end{picture}
\caption{{\it Top panel}: The amount of cold gas ($T<500$ K and
$n_{\text{H}}>500$ cm$^{-3}$) in star forming halos as a
function of redshift.  Star-forming halos are plotted at each
simulation snapshot (separated in time by $3\times 10^6$ years).
{\it Bottom panel}: The total number of star forming halos in
our simulation volume as a function of redshift.}
\end{figure}

\clearpage
\begin{figure}[htb]
\figurenum{2}
\setlength{\unitlength}{1in}
\begin{picture}(6,6.5)
\put(-0.50,-1.8){\includegraphics{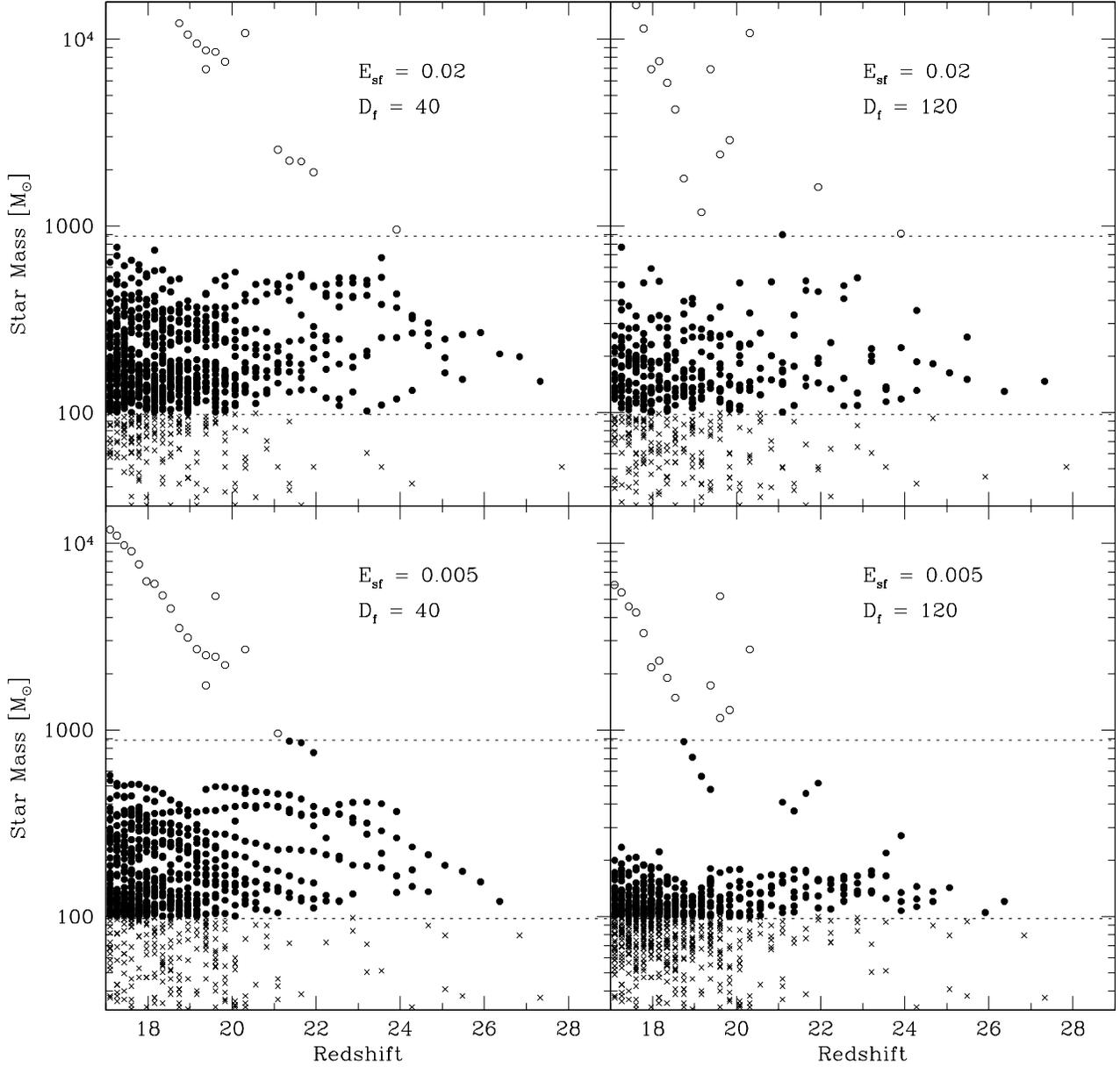}}
\end{picture}
\caption{Total star masses in halos as a function of redshift for
different values of the disruption factor $D_f$ and star formation
efficiency $E_{\rm SF}$ (see text for discussion).  Horizontal
dotted lines indicate the range of allowed star masses in halos
({\it solid-circles}).  Source masses falling
below $100$ M$_{\odot}$ mass ({\it crosses}) are excluded,
while those above $900$ M$_{\odot}$ 
({\it open-circles}) have star masses artificially set 
to an upper limit of $900$
M$_{\odot}$.}
\end{figure}

\clearpage

\begin{figure}[htb]
\figurenum{3} \setlength{\unitlength}{1in}
\begin{picture}(6,6.5)
\put(-0.50,-1.9){\includegraphics{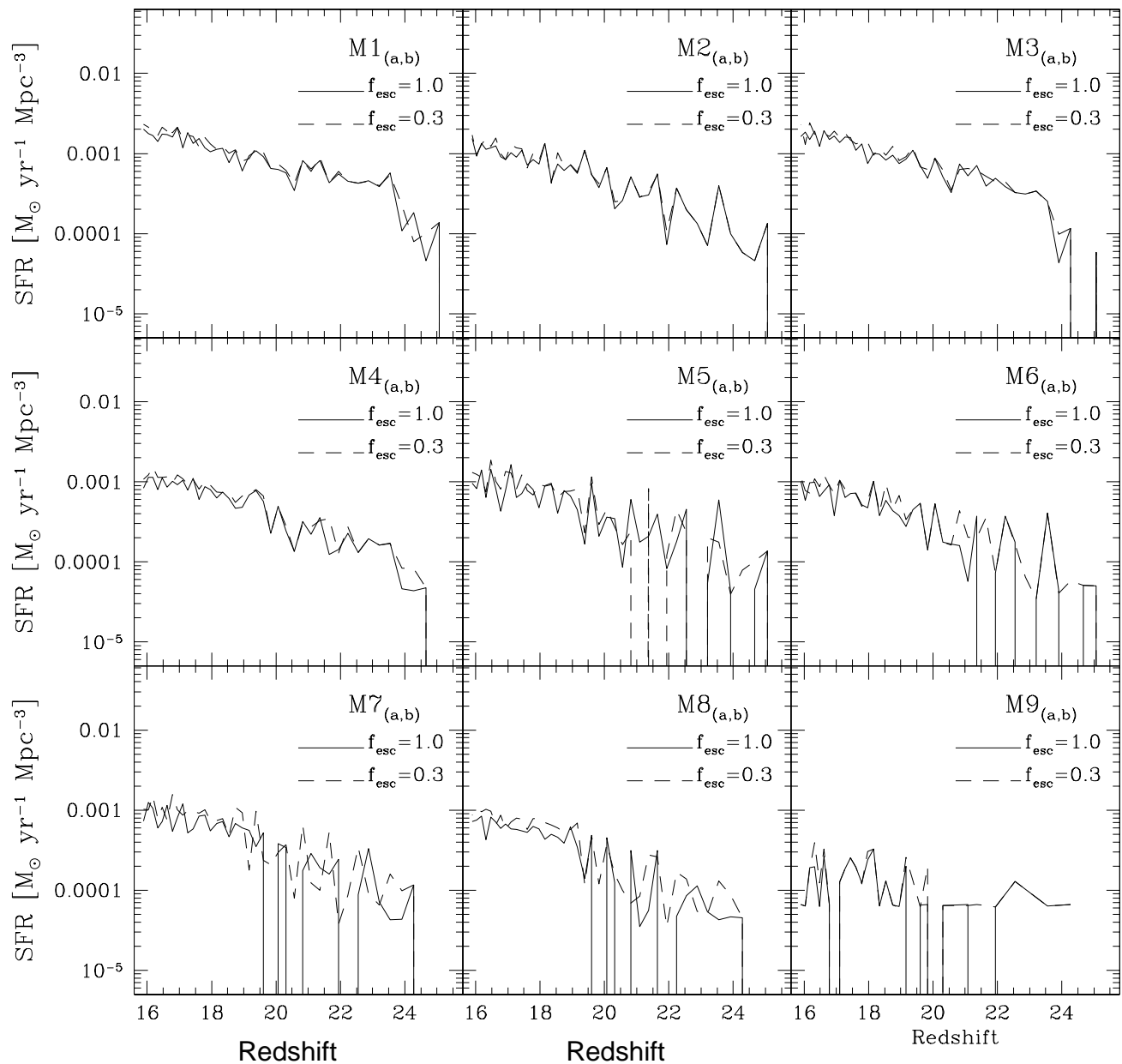}}
\end{picture}
\caption{Evolution of the star formation rate density in each model
down to $z=16$.}
\end{figure}

\clearpage

\begin{figure}[htb]
\figurenum{4}
\setlength{\unitlength}{1in}
\begin{picture}(6,6.5)
\put(-0.50,-1.9){\includegraphics{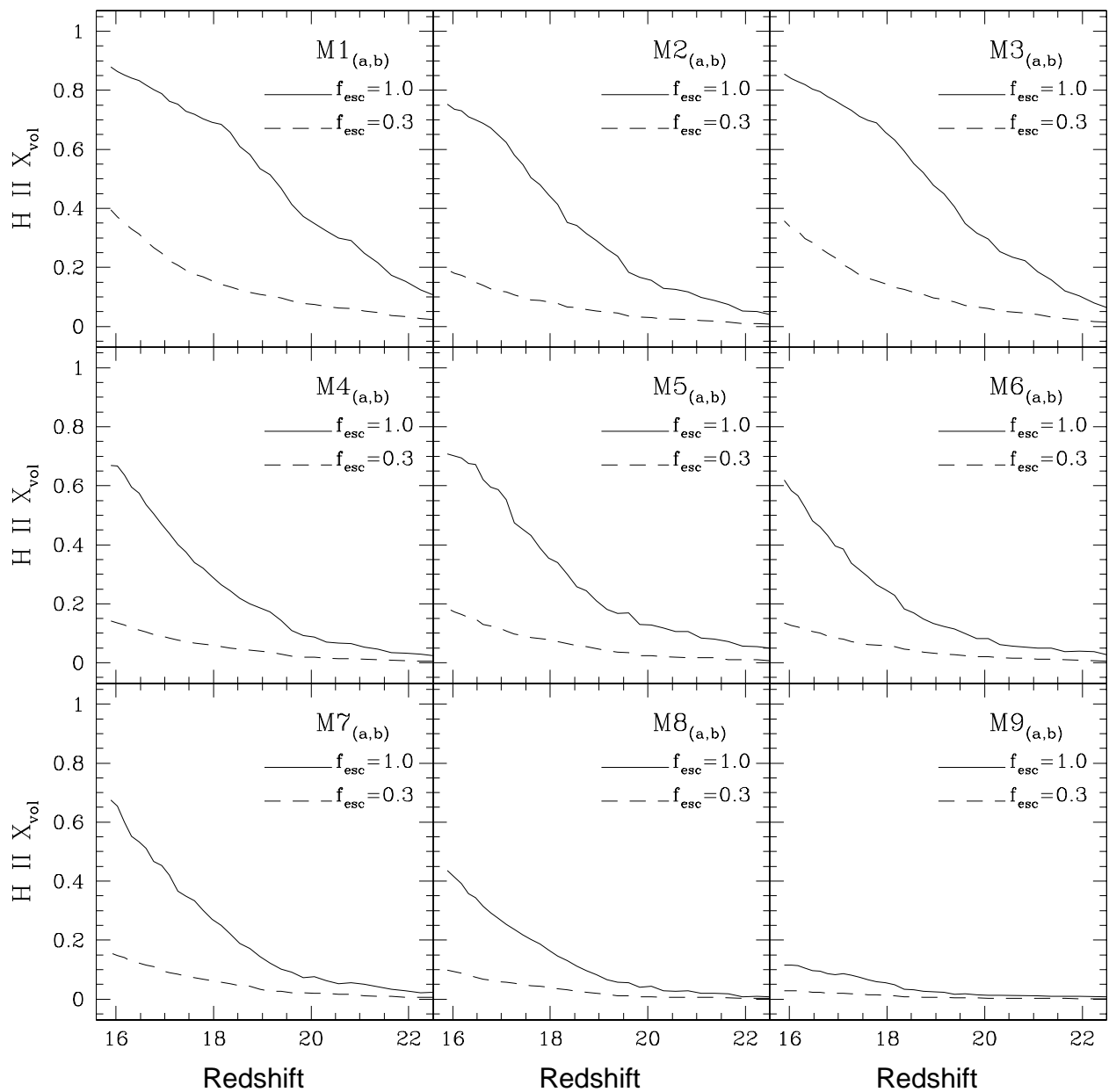}}
\end{picture}
\caption{Evolution of the volume-weighted ionisation fraction for H
{\small II} in each model down to $z=16$.}
\end{figure}

\clearpage

\begin{figure}[htb]
\figurenum{5}
\setlength{\unitlength}{1in}
\begin{picture}(6,6.5)
\put(-0.50,-1.9){\includegraphics{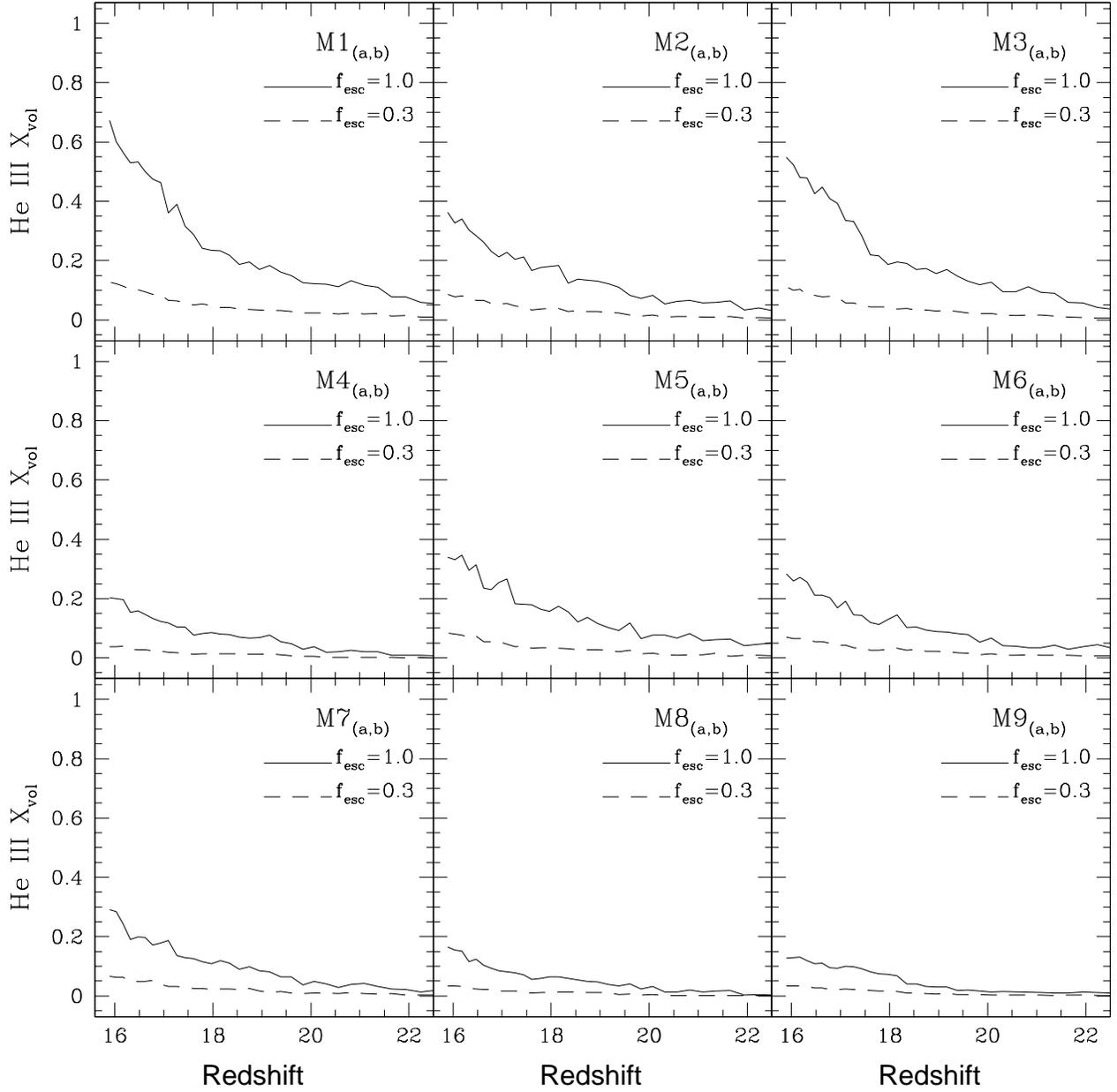}}
\end{picture}
\caption{Same as Figure 4, but for He III.}
\end{figure}

\clearpage

\begin{figure}[htb]
\figurenum{6}
\setlength{\unitlength}{1in}
\begin{picture}(6,7.5)
\put(-0.50,-1.1){\includegraphics{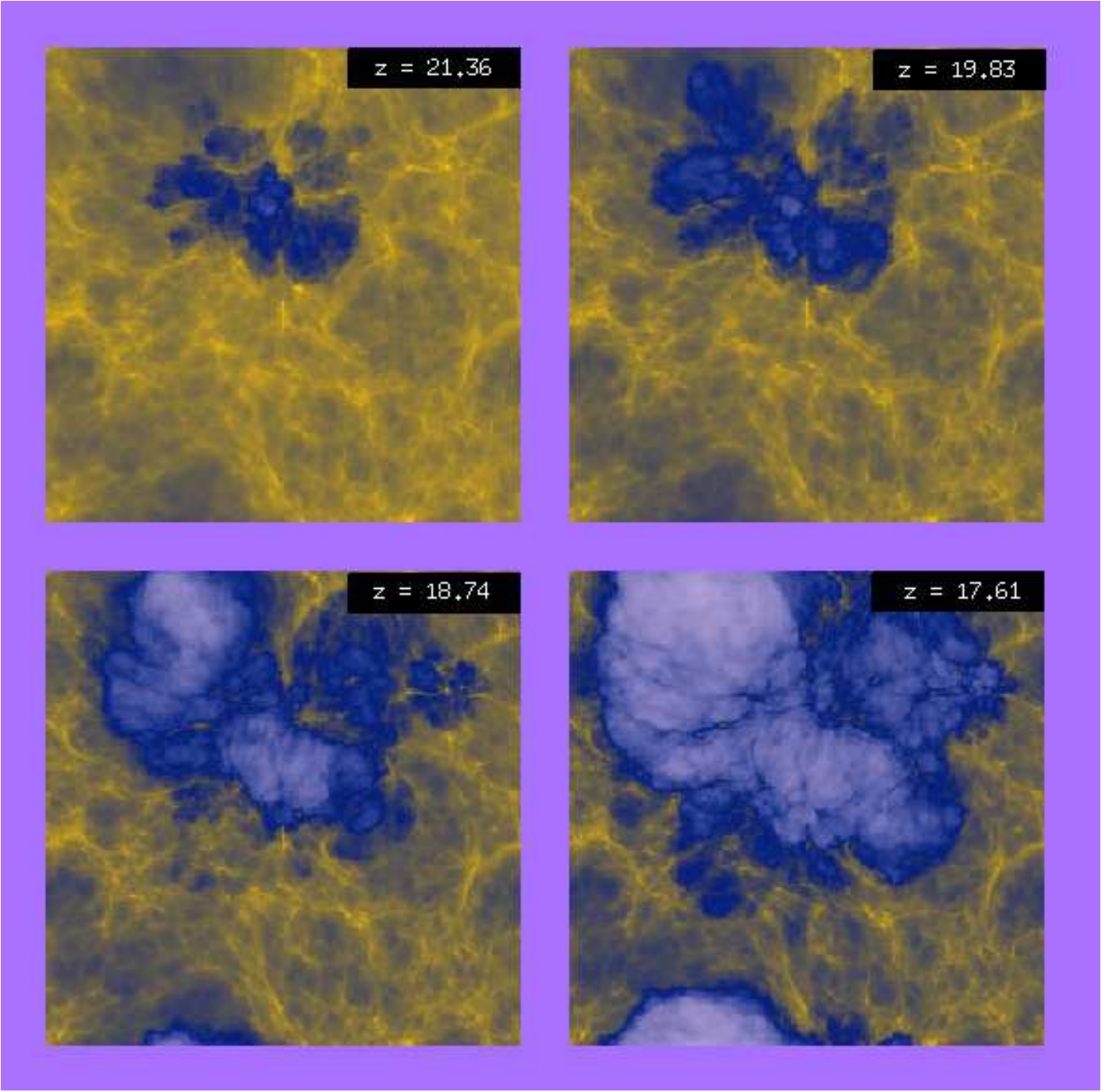}}
\end{picture}
\caption{A series of projected slices through the simulation volume at
({\it top-left} to {\it bottom-right}) z=21.4, 19.8, 18.7, and
17.6.  In each panel, a $0.25$ Mpc slice (1/4 of the box length) from
the outputs of the M1$_{\rm a}$ ($f_{\rm esc}=1.0)$ model is projected in both
density and ionisation fraction.  From the plots one can follow the
growth of the ionisation zones ({\it blue}) around the first stars as
they turn neutral gas ({\it yellow}) into highly ionised regions
({\it light blue}).}
\end{figure}

\clearpage
\begin{figure}[htb]
\figurenum{7}
\setlength{\unitlength}{1in}
\begin{picture}(6,6.5)
\put(0.00,-1.0){\includegraphics{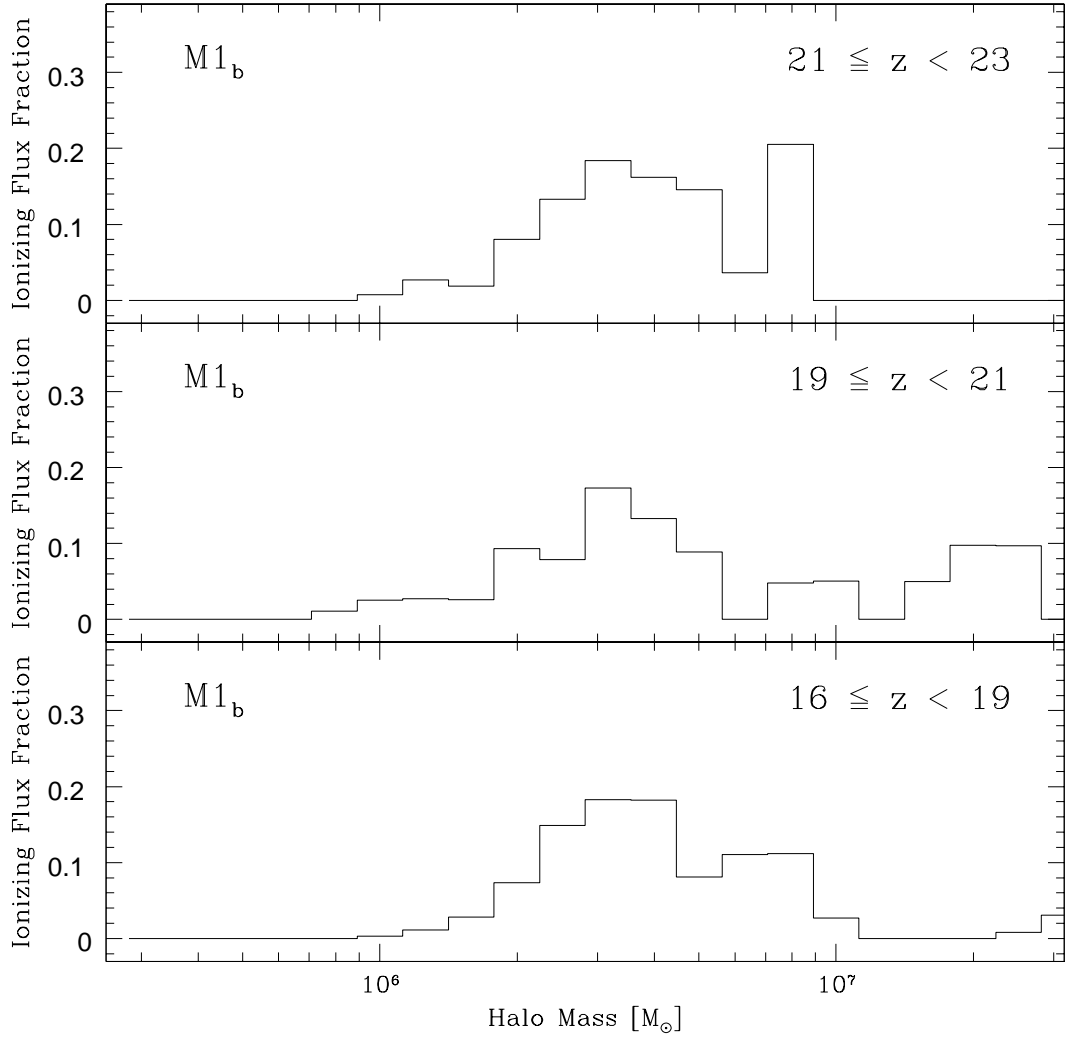}}
\end{picture}
\caption{The fraction of total ionising flux released by population
III stars in model M1$_{b}$ as a function of halo mass for redshift
ranges $21\leq z<23$, $19\leq z<21$, and $16\leq z<19$. Similar
results were obtained for models M2-M9.}
\end{figure}

\clearpage

\begin{figure}[htb]
\figurenum{8}
\setlength{\unitlength}{1in}
\begin{picture}(6,6.5)
\put(-0.50,-1.9){\includegraphics{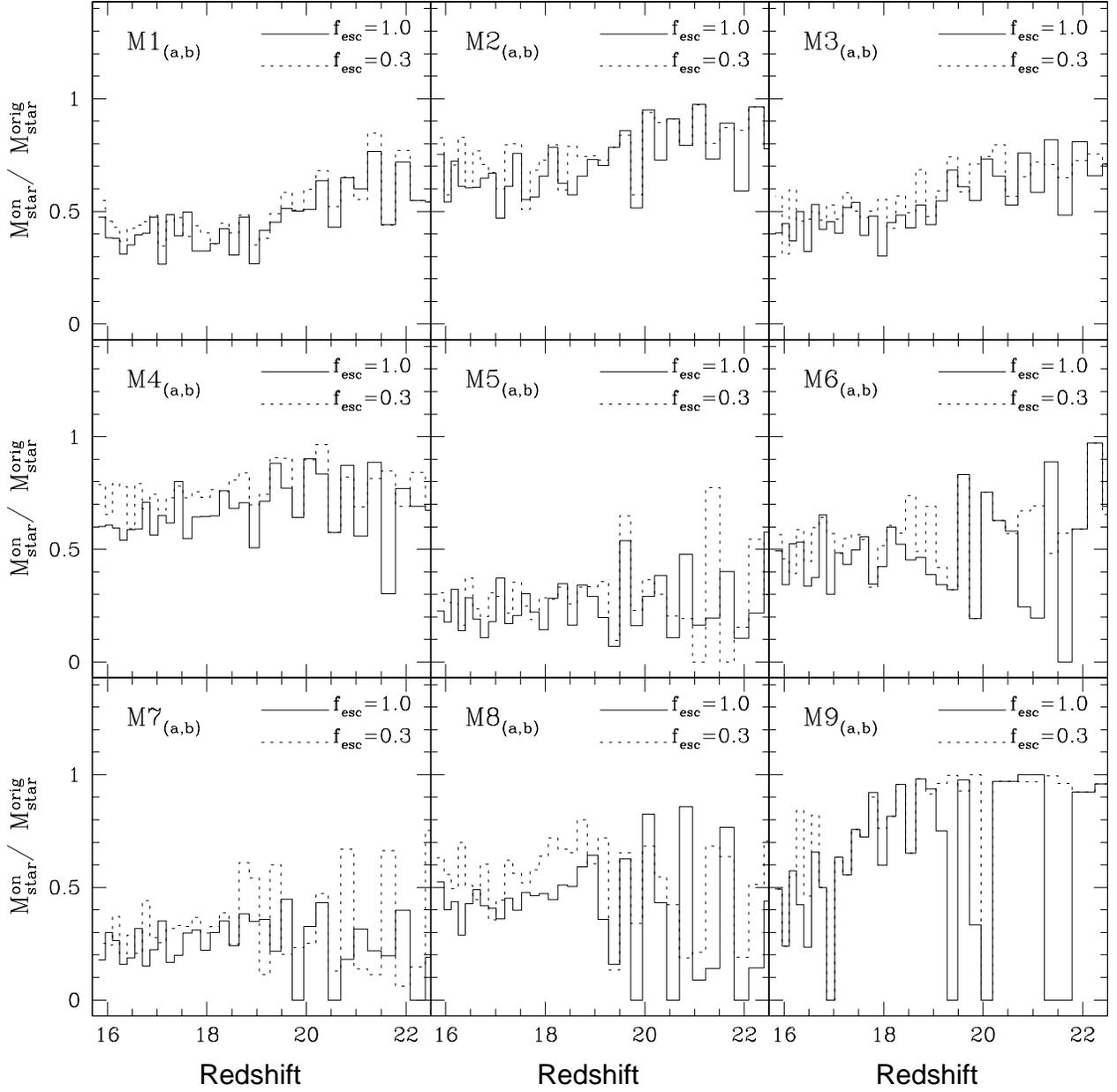}}
\end{picture}
\caption{The evolution of the ratio of the star mass that actually
turns on at each time step $M_{\rm star}^{\rm on}$ to the star mass originally
available $M_{\rm star}^{\rm orig}$ before the effects of the LW background
and the ``ionisation exclusion effect'' were taken into account (see
Table 3 for related statistics).} 
\end{figure}

\clearpage
\begin{figure}[htb]
\figurenum{9}
\setlength{\unitlength}{1in}
\begin{picture}(6,5.3)
\put(0.00,-2.0){\includegraphics{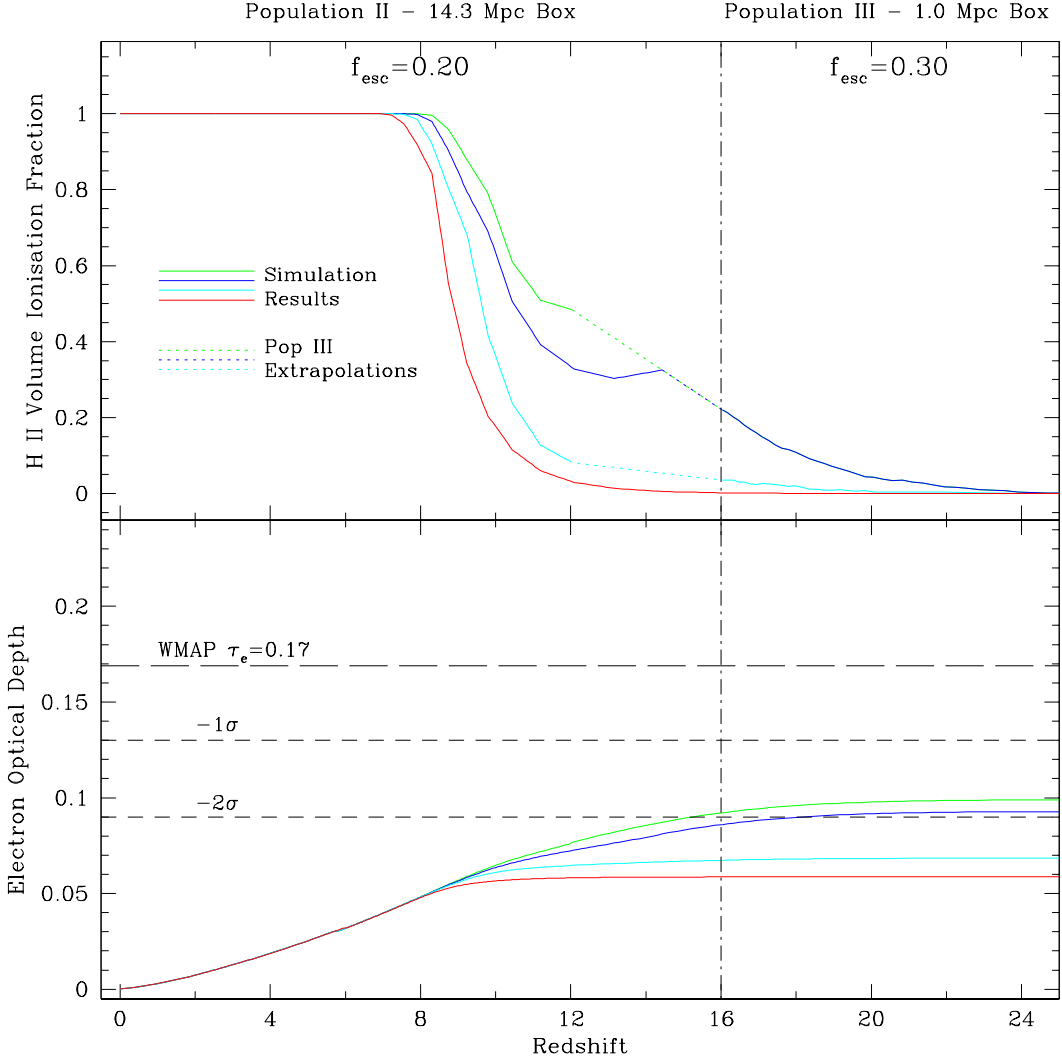}}
\end{picture}
\caption{{\it Top-panel:} The evolution of the H {\small II}
volume-weighted ionisation fraction obtained by linking together the
ionisation histories of $f_{\rm esc}=0.30$ population III simulations
with $f_{\rm esc}=0.20$ population II simulations described in Paper
I. Here the vertical dashed-dotted line represents the redshift
($z=16$) where we were forced to stop our population III simulations
owing to the small box size.  The dotted lines extending below this
redshift represent extrapolations of the ionisation fraction if the
population III epoch had continued to lower redshifts (see text for
discussion). The models shown are (from top to bottom): (1) {\it blue
line}: the combined (mean) population III simulation results from
models M1a-M8a ($f_{\rm esc}=0.30$) extrapolated down to $z=14.5$
before switching to the ($f_{\rm esc}=0.20$) population II simulation,
(2) {\it green line}: same as (1) but with the population III epoch
extrapolated down to $z=12$. (3) {\it cyan line}: model M9a ({\it
ultra-restrictive} ``one-star-per-halo'' model) with a population III
epoch extrapolated to $z=12$ before switching to the population II
simulation, and (4) ({\it red line}): the case where {\it only}
population II sources are present (representing the $f_{\rm esc}=0.20$
model from Paper I).  {\it Bottom-panel:} Corresponding optical depths
to Thomson scattering $\tau_e$ between the present and redshift
$z$. The inferred value for $\tau_e$ from the WMAP measurement is
shown ({\it long-dashed line}) as is the corresponding $1\sigma$ and
$2\sigma$ lower bounds ({\it short-dashed lines}).}

\end{figure}
\clearpage
\begin{figure}[htb]
\figurenum{10}
\setlength{\unitlength}{1in}
\begin{picture}(6,6.5)
\put(0.00,-1.0){\includegraphics{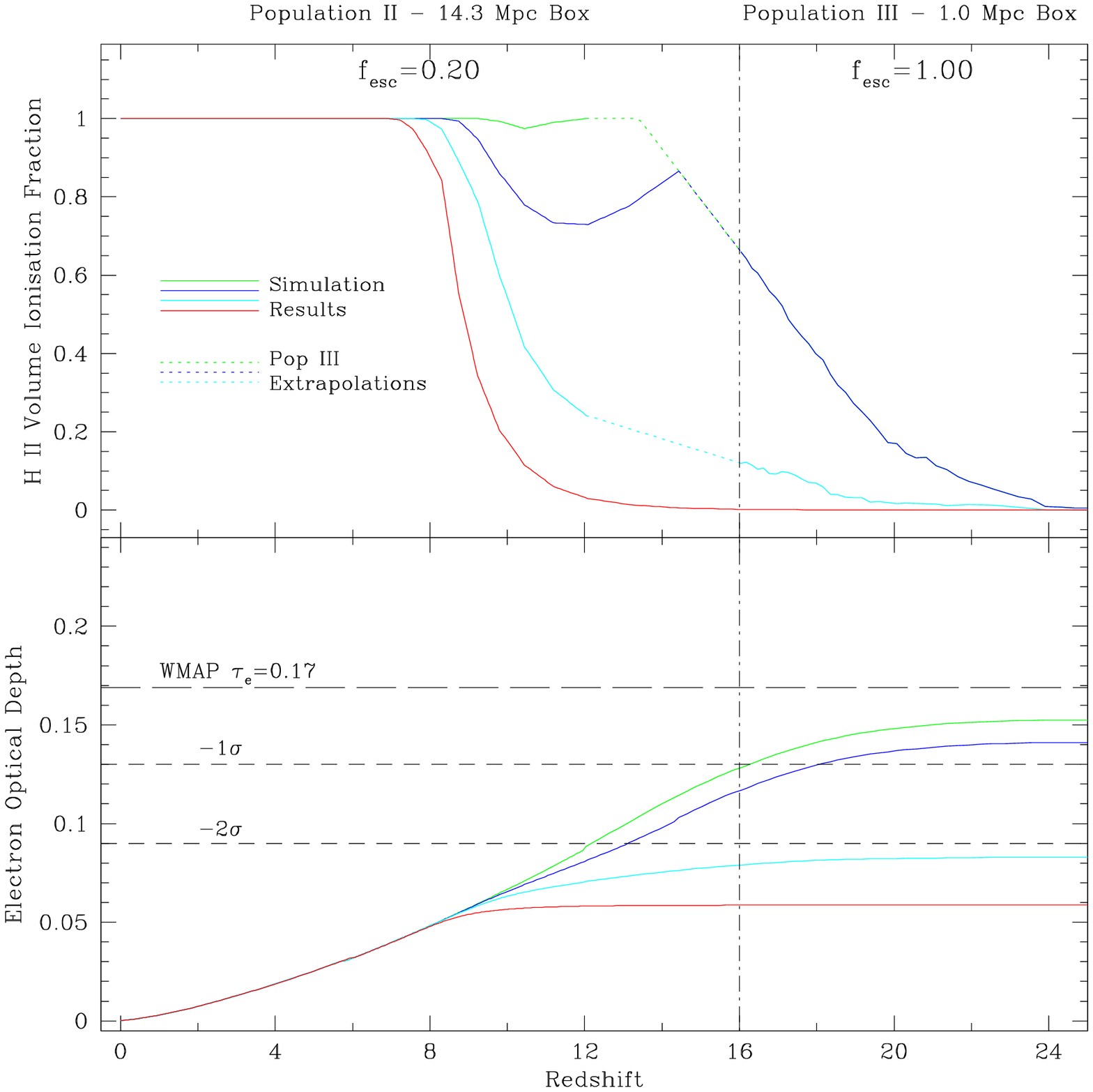}}
\end{picture}
\caption{Same as Figure 9 but for population III sources with $f_{\rm esc}=1.0$.}
\end{figure}


\clearpage

\begin{references}

Abel T., Anninos P., Norman M.L., Zhang Y., 1997, NewA, 2, 181

Abel,T., Anninos, P., Norman, M.~L., \& Zhang, Y.\ 1998, ApJ, 508, 518

Abel T., Bryan G.L, \& Norman M.L., 2000, ApJ, 540, 39

Abel T., Bryan G.L, \& Norman M.L., 2002, Science, 295, 93

Abel T. \& Wandelt, B.D. 2002, MNRAS, 330. L53

Barkana R. \& Loeb A., 2001, Phys. Rep., 349, 125

Barton, E.J., Dav\'e, R., Smith, J.D.T., Papovich, C., Hernquist, L.,

Becker R.H., et al., 2001, AJ, 122, 2850 

Bromm, V., Coppi, P.~S., \& Larson, R.~B.\ 1999, ApJL, 527, L5

Bromm V., Coppi P.S, \& Larson R.B, 2002, ApJ, 280, 825

Bromm V., Ferrara, A., Coppi P.S, \& Larson R.B, 2001, MNRAS, 328, 969

Bromm V., Kudritzki R.P., Loeb A., 2001, ApJ, 552, 464

Bromm V., Yoshida, N., \& Hernquist, L., 2003, ApJ, 596, L135

Cen, R. 2003a, ApJ, 591, 5

Cen, R. 2003b, ApJ, 591, 12

Ciardi B., Ferrara A., \& Abel T., 2000, ApJ, 533, 594

Ciardi, B., Ferrara, A., Governato, F., \& Jenkins, A.\ 2000, MNRAS,
314, 611 

Ciardi B. \& Madau P., 2003, ApJ, 596, 1

Ciardi B., Stoehr F., White, S.D.M., 2003, MNRAS, 343, 1101

Couchman, H.~M.~P.~\& Rees, M.~J.\ 1986, MNRAS, 221, 53

Dekel A. \& Rees M.J., 1987, Nature, 326, 455

Djorgovski S.G., Castro S.M., Stern D., Mahaba A., 2001, ApJ, 560, 5

Fan X., et al. 2000, AJ, 120, 1167

Fan X., et al. 2003, AJ, 125, 1649

Furlanetto S.R., Sokasian A., Hernquist L., 2003, MNRAS, 347, 187

Galli D. \& Palla F., 1998, A\&A, 335, 403 

Gnedin N.Y., 2000, ApJ, 535, 530

Haiman Z., Abel T., \& Rees. M.J., 2000, ApJ, 534, 11

Haiman Z., Rees M.J., \& Loeb A., 1997, ApJ, 476, 458

Haiman Z., Rees M.J., \& Loeb A., 1996, ApJ, 467, 522

Heckman T.M., Sembach K.R., Meurer G.R., Leitherer C., Calzetti D,
Martin C.L., 2001, ApJ, 558, 56

Heger A. \& Woosley S.E. 2002, ApJ, 567, 532

Hernquist L. \& Springel V., 2003, MNRAS, 341, 1253

Hui L. \& Haiman Z., 2003, ApJ, 596, 9

Hurwitz M., Jenlinsky P., Dixon W. V., 1997, 498, L31

Jang-Condell, K. \& Hernquist, L., 2001, ApJ, 548, 68

Kitayama, T., Susa, H., Umemura, M. \& Ikeuchi, S. 2001, MNRAS, 326, 1353

Kogut A., et al. 2003, ApJ, 148, 161

Leitherer C., Ferguson H.C., Heckman T.M., Lowenthal J.D., 1995, ApJ
454, L19

Machacek M.E., Bryan G.L., \& Abel T., 2003, MNRAS, 338, 273

Machacek M.E., Bryan G.L., \& Abel T., 2001, ApJ, 548, 509

Mackey J., Bromm V., \& Hernquist L., 2003, ApJ, 586, 1

Oh S.P., 2001, ApJ, 569, 558

Oh S.P., Haiman, Z., \& Rees, M. J., 2001, ApJ, 553, 73

Oh S.P., Nollett K.P., Madau P., Wasserburg G.J., 2001, ApJ, 562L, 1

Omukai K., 2000, ApJ, 534, 809

Omukai K. \& Nishi R., 1999, ApJ, 518, 64

Razoumov A.O., Norman M.L., Abel T., Scott D., 2002, ApJ, 572, 695

Ricotti M., Gnedin N.Y., \& Shull J.M., 2002a, ApJ, 575, 49
 
Ricotti M., Gnedin N.Y., \& Shull J.M., 2002b, ApJ, 575, 33

Ricotti M. \& Shull J.M, 2000, ApJ, 542, 548 

Salvaterra, R., Ferrara, A., \& Schneider, R., 2003, [astro-ph/0304074]

Schaerer D., 2002, A\&A, 382,28

Schneider, R., Ferrara, A., Natarajan, P. \& Omukai, K., 2002, 
ApJ, 571, 30

Sokasian A., Abel T., Hernquist L., Springel V., 2003, MNRAS, 344, 607
(Paper I)

Sokasian A., Abel T., \& Hernquist, 2002, MNRAS, 332, 601

Sokasian A., Abel T., \& Hernquist, 2001, NewA, 6, 359

Spergel, D.N. et al. 2003, ApJ, 148, 195

Springel V. \&  Hernquist L., 2003, MNRAS, 339, 312

Springel V. \&  Hernquist L., 2002, MNRAS, 333, 649

Springel, V., 2003, ApJ, submitted [astro-ph/0310514]

Steidel C.C., Pettini M., Adelberger K. L., 2001, ApJ, 546, 665

Theuns T., Schaye J., Zaroubi S., Kim T., Tzanavaris P., Carswell B.,
2002, ApJ, 567, L103

Tumlinson J., Giroux M.L., \& Shull M.J., 2001, ApJ, 550, L1 

Tumlinson J. \& Shull J.M., 2000, ApJ, 528, L65

Venkatesan A., Tumlinson J., Shull M.J., 2003, ApJ, 584, 621

Whalen D., Abel T., \& Norman M.L., 2003, ApJ, submitted [astro-ph/0310283]

White R.L., Becker R.H., Fan X., Strauss M.A., 2003, ApJ, 126, 1

Wood  K., Loeb A., 2000, ApJ, 545, 86

Wyithe J.S.B. \& Loeb A., 2003a, ApJ, 588, 69

Wyithe J.S.B. \& Loeb A., 2003b, ApJ, 586, 693

Yoshida, N., Sokasian A., Hernquist, L., Springel V., 2003a, ApJ, 598, 73

Yoshida, N., Sokasian A., Hernquist, L., Springel V., 2003b, ApJ, 591, 1

Yoshida, N., Abel, T., Hernquist, L., Sugiyama, N., 2003c, ApJ, 592, 645

Yoshida, N., Sugiyama, N., Hernquist, L., 2003d, MNRAS, 344 481

Yoshida, N., Bromm, V., Hernquist, L., 2003e, ApJ, submitted
[astro-ph/0310443]

Zaldarriaga, M., Furlanetto, S.R., Hernquist, L., 2003, ApJ, submitted

\end{references}
\end{document}